# Recovering Nonlinear Functions of Latent Variables: A Plausible-Value Neural Network Framework

Eunjeong Song · Sehee Hong
Department of Education, Korea University, Seoul, Republic of Korea

Eunjeong Song, ORCID iD: https://orcid.org/0000-0002-2302-3227
Sehee Hong, ORCID iD: https://orcid.org/0000-0001-5468-8398

## Abstract

When factor scores replace true latent scores in nonlinear prediction, measurement error attenuates the recoverable variance of any $k$th-order component of the regression function by $\rho^k$—the $k$th power of the score's coefficient of determination—for any linear score type. This study derives the bound via Hermite polynomial expansion and proposes PV-ANN—plausible values (posterior draws preserving latent variance) combined with artificial neural networks (learning functional form without prespecification). The bound governs recovery of the latent-scale function, not prediction of the outcome from observed indicators, for which factor scores are already sufficient; the two metrics are therefore predicted to dissociate. An 18-condition simulation supports both predictions: in the nonlinear low-reliability conditions PV-ANN closes about four fifths of the function-shape recovery gap between a factor-score learner and one given the true latent values, and the margin widens as reliability falls, while predictive accuracy is not improved, as the theory requires. A Big Five application illustrates the intended exploratory workflow and delineates boundary conditions under weak signal and measurement model misspecification.

## 1 Introduction

When error-laden scores stand in for latent variables in a prediction model, the estimated relationship is systematically distorted. In the linear case this is the classical attenuation of regression coefficients (Carroll et al. 2006; Fuller 1987), documented for test scores entered as predictors (Bhaktha and Lechner 2021) and, more generally, for structural parameters estimated from predicted latent scores (Croon 2002); comparable effects have been reported in network psychometric models (de Ron et al. 2022). In the nonlinear case the distortion is qualitatively different: the shape of the function is flattened, so nonlinearity can escape detection altogether (Section 1.1).

This study derives the theoretical mechanism of nonlinear smoothing ($\rho^k$ attenuation), proposes the PV-ANN framework—combining plausible values (PVs) with artificial neural networks (ANNs)—as a tool to address it, and quantitatively tests the theoretical predictions through simulation. PVs integrate measurement uncertainty via posterior distribution sampling, and ANNs learn the shape of nonlinear relationships without prior specification.

In the vocabulary of statistical learning this is an errors-in-variables problem in which the noise model is not unknown but is identified up to a parametric factor model. The question the paper answers is what that structure buys: posterior sampling strictly dominates plug-in point estimates for recovering the regression function on the latent scale, and cannot improve on them for predicting the outcome from the observed indicators. The two are separate estimands, and the separation is what makes the framework's scope statable in advance rather than discovered by benchmarking.

### 1.1 Factor Score Shrinkage and Nonlinear Signal Loss

When a latent variable $\eta$ is measured by congeneric indicators (Jöreskog 1971), the regression-based factor score is expressed as

$$\tilde{\eta} = \rho\eta + \sqrt{\rho(1-\rho)}\,\epsilon^* \qquad (1)$$

(Skrondal and Laake 2001), where $\epsilon^* \sim N(0,1)$ and $\epsilon^* \perp \eta$, so $\mathrm{Var}(\tilde{\eta}) = \rho$.

$$\rho = \frac{\boldsymbol{\lambda}'\boldsymbol{\Psi}^{-1}\boldsymbol{\lambda}}{1 + \boldsymbol{\lambda}'\boldsymbol{\Psi}^{-1}\boldsymbol{\lambda}} \qquad (2)$$

is the coefficient of determination between the latent variable and the factor score, where $\boldsymbol{\Psi} = \mathrm{diag}(\psi_1, \ldots, \psi_J)$. $\rho$ is distinct from scale-level composite reliability $\omega$ (McDonald 1999), which is used hereafter as the label for measurement conditions; in congeneric models, regression scores use indicator-specific optimal weights so that generally $\rho \geq \omega$. Equation (1) is exact in the single-factor case; for correlated multivariate factors, the scalar $\rho$ is replaced by diagonal elements of the coefficient of determination matrix (Skrondal and Laake 2001, Eq. 13). In the present simulation with latent correlation .30, the difference between univariate $\rho$ and the multivariate diagonal was within .004 (Appendix C.3).

If the true relationship is $f(\eta) = \eta^2$, the optimal prediction function from a model using factor scores is $E[\eta^2 \mid \tilde{\eta}] = \tilde{\eta}^2 + (1-\rho)$: the functional form remains quadratic, but the variance of the conditional expectation relative to that of $\eta^2$ is reduced to $\rho^2$. More generally, in the single-factor setting, expanding $f(\eta)$ in Hermite polynomials gives $f(\eta) = \sum_{k=0}^{\infty} c_k\, H_k(\eta)$, and the capturable variance ratio of the $k$th-order component decreases to $\rho^k$ (Appendix A.1). Hereafter, *nonlinear signal* refers to the aggregate contribution of all $k \geq 2$ Hermite components. The case $k = 2$ is the classical result that the reliability of a squared or product term is the product of the component reliabilities (Busemeyer and Jones 1983); Equation (A3) generalizes it to arbitrary order.

**The bound does not depend on the choice of score.** This $\rho^k$ ceiling is a property of the information a score carries, not of the particular score. For any linear composite $s$ of the indicators, the capturable variance ratio of the $k$th-order component is $\rho_s^k$, where $\rho_s$ is that score's own coefficient of determination (Appendix A.2). Regression and Bartlett scores share the same $\rho$ because both are invertible linear transformations of the jointly sufficient statistic $\boldsymbol{\Lambda}'\boldsymbol{\Psi}^{-1}\boldsymbol{X}$, so that $\sigma(\tilde{\eta}) = \sigma(\tilde{\eta}_B)$; the sum score, whose $\rho_s$ equals $\omega$ under the present design, is bounded more tightly still. The bound is tested empirically via the Bart-ANN benchmark (specified in Section 3.4; results in Sections 4.3 and 6.1). Independent evidence outside psychometrics is reported in Spicker et al. (2025), who demonstrated that measurement error in dietary intake data degrades neural network prediction below misspecified linear baselines.

**What the bound is a limit on.** Two distinct statements follow, and the study depends on keeping them apart. For *prediction*, the bound is a processing limit rather than a score-specific artifact: conditional expectation given the score is an $L^2$ contraction on the Hermite basis (Appendix A.1), so no downstream transformation of $\tilde{\eta}$ can increase the share of $f(\eta)$ that is predictable from the score (cf. the data processing inequality; Cover and Thomas 2006). For *recovery of $f$ on the latent scale*, the map $c_k \mapsto c_k\rho^{k/2}$ is invertible whenever $\rho$ is known, so $\rho^k$ attenuation states what the plug-in estimator loses, not what the data fail to identify. Inverting it is a deconvolution, which converges only at logarithmic rates under normal measurement error (Fan and Truong 1993); PV-ANN reaches the same estimand through the posterior instead, which is why a posterior-sampling framework is proposed here rather than a moment correction. What buys the faster rate is the factor model: the posterior $p(\boldsymbol{\eta} \mid \boldsymbol{X})$ supplies the inversion parametrically rather than nonparametrically, so the recovery advantage is only as good as that model. The condition on measurement specification in Section 6.1 is therefore a consequence of the mechanism, not an incidental caveat. Extensions to non-Gaussian measurement models are discussed in Section 6.3.3.

**Mechanism of loss recovery by PVs.** PVs are random draws from the posterior $p(\eta \mid \boldsymbol{X}, Y)$, so the original latent variance is preserved. Given $Y_i = f(\boldsymbol{\eta}_i) + \epsilon_i$ with $\epsilon_i \perp (\boldsymbol{\eta}_i, \boldsymbol{x}_{\text{obs},i})$, the ensemble average of $M$ predictors trained on $M$ PVs approximates

$$\frac{1}{M}\sum_{m=1}^{M} \hat{f}^{(m)}\left(\boldsymbol{\eta}_j^{(m,\text{train})}\right) \approx E[f(\boldsymbol{\eta}_j) \mid \boldsymbol{x}_{\text{obs},j}, Y_j] \qquad (3)$$

at the training stage where $Y_j$ is observed; the prediction-stage form is Equation (4) below. The right-hand side is the posterior expectation of $f$ across all orders, so $\rho^k$ attenuation is in principle eliminated. Numerically, at $\rho \approx .818$ ($\omega = .80$), only about 67% of the second-order signal is accessible via factor-score-based prediction; at $\rho \approx .949$ ($\omega = .94$), 90%.

In the single-factor setting, the absolute advantage conferred by PVs is $1 - \rho$ for the linear case and $1 - \rho^k$ for the $k$th-order case, so the ratio of the nonlinear to linear advantage is $1 + \rho + \cdots + \rho^{k-1}$ (Appendix A.3). This ratio monotonically increases with the order of nonlinearity, so the advantage of PVs is amplified in nonlinear prediction. This theoretical prediction provides the basis for RQ 3.

**Remark 1 (two estimands, two predictions).** The $\rho^k$ bound concerns recovery of $f$ on the latent scale, not prediction of $Y$ from observed indicators. For the latter, the optimal predictor is $E[Y \mid \boldsymbol{x}_{\text{obs}}] = E[f(\boldsymbol{\eta}) \mid \boldsymbol{x}_{\text{obs}}]$, which—because regression and Bartlett scores are invertible transformations of the jointly sufficient statistic $\boldsymbol{\Lambda}'\boldsymbol{\Psi}^{-1}\boldsymbol{X}$ (Appendix A.2)—is a function of the factor score alone. A point-estimate-based learner can therefore attain the predictive optimum, whereas the PV ensemble (Equation (4) in Section 2.1) approximates the *same* conditional expectation while adding finite-$M$ between-imputation variance. Under a correctly specified Gaussian measurement model, the theory thus makes two distinct predictions: (i) on function shape recovery, PVs dominate point estimates, with the margin governed by $1 - \rho^k$; (ii) on predictive accuracy for $Y$ (RMSE and cross-validated $R^2$), no PV advantage is possible in principle, and a small deficit of order $O(1/M)$ is expected. Both predictions are evaluated in Study 1, and prediction (ii) supplies the evaluation logic of Study 2 (Section 5.3). When the measurement model is misspecified, the sufficiency premise weakens; this motivates boundary condition (c) in Section 6.1.

### 1.2 Existing Approaches and the Position of the Present Study

**Parametric confirmatory approaches.** LMS (latent moderated structural equations; Klein and Moosbrugger 2000), structural after measurement (SAM; Rosseel et al. 2025; Rosseel and Loh 2024), the two-stage method of moments (Wall and Amemiya 2003), product-indicator approaches (Marsh et al. 2004), and bias-corrected factor score regression (Devlieger et al. 2016; see also Bogaert et al. 2023, 2026; Devlieger et al. 2019) estimate nonlinear effects while controlling for measurement error but require prior specification of the relationship form. LMS is included in the present simulation as a benchmark under correct specification and misspecification (Section 3.4).

**Nonparametric point-estimate approaches.** Grønneberg and Irmer (2024) entered Bartlett factor scores into nonparametric regression. Because factor scores are point estimates, the $\rho^k$ signal loss in Section 1.1 applies. The statistical counterpart is errors-in-variables nonparametric regression, which recovers the regression function by deconvolution (Section 1.1); the factor model avoids deconvolution by supplying the posterior of $\boldsymbol{\eta}$ directly. That literature also separates the two estimands treated here, having established that predicting $Y$ does not require estimating the regression function itself (Carroll et al. 2009; cf. Remark 1).

**Joint optimization approaches.** LatentNN (Ting 2026; originally developed for astronomical spectral analysis) simultaneously optimizes true input values and network parameters. It does not leverage the factor model structure, presupposes specification of input error variance, and uses point optimization rather than posterior sampling. Direct comparison is deferred to future research (Section 6.3.3).

**Machine learning–psychometrics integration.** The present work joins a growing effort to couple machine learning with psychometric measurement models (Templin 2025), in which neural architectures have been paired with latent variable models for estimation and scoring (Couto Tabak et al. 2025; Gonzalez 2025) and with likelihood-intractable process models for inferring time-varying latent states (Pan et al. 2025). Within this effort, measurement error has been identified as an underexamined threat to machine-learning findings in psychology, capable of masking true nonlinear relationships regardless of sample size (Jacobucci and Grimm 2020). PV-ANN addresses that side of the integration: it formalizes when point-estimate inputs cap the nonlinear information available to a flexible learner and how posterior sampling restores it.

**PV-based linear approaches.** Schofield et al. (2015) proposed using PVs as predictors, and Bhaktha and Lechner (2021) documented satisfactory PV performance in linear regression. This approach is limited to linear structural models.

Within the broader landscape of stepwise latent variable estimation (Vermunt 2025), PV-ANN extends the measurement error control that bias-corrected factor score regression achieved in linear models to exploratory contexts in which the form of the nonlinear relationship is unknown.

**Contribution.** The study makes three contributions. First, it derives the $\rho_s^k$ attenuation law for any linear score (Appendix A.1, A.2), establishing that the nonlinear recovery limit of point-estimate-based prediction depends on the score only through its own coefficient of determination and not on the choice of score type. Second, it proposes PV-ANN, which combines posterior-sampling-based integration of measurement uncertainty with form-free nonlinear learning, and formalizes when the framework can and cannot help (Remark 1). Third, it tests the resulting predictions in an 18-condition simulation and probes their boundary conditions in field data, yielding a two-stage workflow in which PV-ANN identifies functional form in the exploration stage and a parametric model (e.g., LMS) provides confirmatory estimation.

### 1.3 Research Questions

The $\rho^k$ theory (Section 1.1) and Remark 1 yield three testable predictions, formalized as RQ 1–3 below. RQ 4 addresses the domain question of whether the resulting advantage transfers to field data; the conditions delimiting this domain—one scaling condition and two candidate boundary conditions—are developed in Section 6.1 on the basis of the combined evidence.

**RQ 1.** Does entering PVs into an ANN improve nonlinear function shape recovery relative to point-estimate-based entry (sum score, regression factor score, Bartlett factor score)? The primary evaluation metric is function shape recovery ($R^2_{\text{recovery}}$; Section 3.6), which quantifies the goodness of fit of the estimated function to the true function on the latent variable grid. Because PV posterior sampling entails between-imputation variance, a trade-off with point-estimate-based approaches arises in individual prediction accuracy (root mean square error, RMSE), which is used as a supplementary metric.

**RQ 2.** Does this improvement increase as measurement reliability decreases? Following the $\rho^k$ attenuation argument, the recovery margin for PVs expands at lower reliability.

**RQ 3.** Is the PV advantage larger when coupled with a nonlinear (ANN) predictor than with a linear (LR) predictor, and does this within-DGP amplification exceed the baseline observed under the linear DGP? This is a consistency check on the same $\rho^k$ law that RQ 1 and RQ 2 test more directly, not an independent test of it (Section 4.5). The PV advantage in the linear case has been established in prior research (Bhaktha and Lechner 2021; Schofield et al. 2015); RQ 3 asks whether the nonlinear combination exceeds it. This is operationalized within each DGP as the amplification difference: (PV advantage with nonlinear ANN predictor) − (PV advantage with linear LR predictor). The Diff value under the linear DGP serves as a baseline representing predictor-type amplification independent of DGP nonlinearity; Diff values under nonlinear DGPs in excess of this baseline reflect nonlinear amplification (Section 3.6). Equation (A6) is monotone in the order $k$, but the two nonlinear DGPs here differ in the order of their dominant component *and* in its share of explained variance, so what the design tests is the $k \geq 2$ versus $k = 1$ contrast, not monotonicity in $k$.

**RQ 4.** Do the Study 1 findings transfer to field data whose properties—signal strength, reliability, and measurement model adequacy—depart from the simulation's favorable conditions? Study 2 probes this domain question with a deliberately demanding application ($R^2_{\text{total}} \approx .05$, $\omega \approx .85\text{–}.90$, confirmatory factor analysis (CFA) fit below conventional cutoffs). Because true latent function values are unobservable in field data, RQ 4 is evaluated (a) on the predictive metric, for which Remark 1 predicts parity among PV- and point-estimate-based learners—making cross-validated $R^2$ a specification check of the theory rather than a horse race—and (b) qualitatively via the partial dependence plot, which displays the shape of the estimated function (Section 5.3).

# 2 Methodological Framework

## 2.1 PVs and Predictive Inference

PVs are random draws from the posterior distribution of latent variables given observed data $\boldsymbol{x}_{\text{obs},i}$ (Mislevy 1991; Lüdtke and Robitzsch 2017). The use of PVs here is based on approximating the predictive expectation with respect to measurement uncertainty, not on population-level parameter estimation (von Davier et al. 2009). Under the conditional independence assumption of the measurement model, by the law of iterated expectations,

$$E[Y_i \mid \boldsymbol{x}_{\text{obs},i}] = E_{\boldsymbol{\eta}_i \mid \boldsymbol{x}_{\text{obs},i}}[f(\boldsymbol{\eta}_i)] \approx \frac{1}{M} \sum_{m=1}^{M} \hat{f}^{(m)}\,(\boldsymbol{\eta}_i^{(m)}) \qquad (4)$$

Here, $\hat{f}^{(m)}$ is the ANN trained on the $m$th PV dataset in Step B and is applied without retraining at the prediction stage. The key difference between Equations (3) and (4) is the conditioning scope: $\boldsymbol{\eta}_j^{(m,\text{train})}$ is drawn with $Y$ included, whereas $\boldsymbol{\eta}_i^{(m)}$ is drawn at the prediction stage with the test set $Y_i$ treated as missing.

The approximation rests on two conditions. The first is consistency: each $\hat{f}^{(m)}$ must converge to the true function, for which the universal approximation property of ANNs (Hornik et al. 1989) supplies representability, with convergence under regularization and early stopping assessed empirically by the Oracle ratio of Section 4.3. The second is congeniality between the imputation and analysis models (Meng 1994), which here holds only approximately; Section 4.2 evaluates whether the resulting distortion materially degrades $R^2_{\text{recovery}}$. Schofield et al. (2015) showed that omitting $Y$ from the conditioning model induces bias, a source the present design removes by including it (Section 2.2). Equation (3) extends Rubin's (1987) pooling to a nonlinear predictor.

### 2.2 Inclusive Imputation Model

Following the inclusive imputation strategy (Collins et al. 2001; van Buuren 2018, Ch. 6), the conditioning model of the Gibbs sampler (also referred to as the imputation model in the multiple-imputation literature; the two terms are used interchangeably hereafter) includes $Y$, quadratic terms for all latent variables, and all pairwise interaction terms:

$$Y_i = \gamma_0 + \sum_{p=1}^{P} \gamma_p \, \eta_{pi} + \sum_{p=1}^{P} \gamma_{pp} \, \eta_{pi}^2 + \sum_{p<q} \gamma_{pq} \, \eta_{pi}\eta_{qi} + e_i \qquad (5)$$

Here, $e_i$ is the residual of the conditioning model, distinguished from the structural model error $\epsilon_i$ defined in Section 3.2. Inclusion of $Y$ simultaneously satisfies the inclusiveness principle and the bias prevention condition of Schofield et al. (2015). The quadratic and interaction terms serve as second-order Taylor expansion approximations to an unknown nonlinear DGP. This approximation is exact for DGP 2 (quadratic) and an intentional misspecification for DGP 3 (sigmoid; Section 3.2), the latter functioning as a stress test for practical robustness. The conditioning model is not the analysis model: its role is to generate posterior draws of $\boldsymbol{\eta}$, whereas the form of the relation between $\boldsymbol{\eta}$ and $Y$ is left to the ANN. DGP 3 separates the two, since there the conditioning model is misspecified for the true $f$ while the ANN nonetheless recovers it (Section 4.3).

### 2.3 Three-Stage Protocol for Training and Prediction

**Step A.** The full data are split 80:20 into training and test sets by simple random sampling, with a distinct random seed per replication (Study 1) and per empirical holdout (Study 2). Seeds are not stored as a list: every stream is derived deterministically from a single recorded base seed by the `derive_seeds` function of the archived code, which fixes the data, sampler, and network-initialization seeds of each replication and holdout, and `scripts/export_seeds.py` writes the resulting tree out in full (Data and Code Availability).

**Step B (Training).** The Gibbs sampler is run on the training set with $Y_{\text{train}}$ included in the conditioning model. $M$ posterior parameter sets are drawn at equal intervals from the post-burn-in interval, and the corresponding latent variable draws serve as PVs for the training set. For each PV dataset, $K$ ANNs with different random seeds are trained, and the mean of the $K$ predictions is used (Section 2.4).

**Step C (Prediction).** For each of the $M$ posterior parameter sets, a reduced Gibbs sampler is run for each test observation $i$, holding $\boldsymbol{\theta}^{(m)}$ fixed and sampling only $\boldsymbol{\eta}_i$. The test set $Y_i$ is treated as missing, and two steps are iterated:

Step (i): $Y_i$ is predictively imputed from the current $\boldsymbol{\eta}_i$ estimate:

$$Y_i^{(imp)} \sim N\left(\boldsymbol{\gamma}^{(m)\prime}\boldsymbol{z}(\boldsymbol{\eta}_i^{(t)}),\, \sigma^{2(m)}\right) \qquad (6)$$

where $\boldsymbol{z}(\boldsymbol{\eta}_i)$ is the design vector of the conditioning model.

Step (ii): $\boldsymbol{\eta}_i$ is sampled from its posterior distribution using both the imputed $Y_i^{(imp)}$ and the observed $\boldsymbol{x}_{\text{obs},i}$:

$$p(\boldsymbol{\eta}_i \mid \boldsymbol{x}_{\text{obs},i}, Y_i^{(imp)}, \boldsymbol{\theta}^{(m)}) \propto p(\boldsymbol{x}_{\text{obs},i} \mid \boldsymbol{\eta}_i, \boldsymbol{\Lambda}^{(m)}, \boldsymbol{\Psi}^{(m)})\; p(Y_i^{(imp)} \mid \boldsymbol{\eta}_i, \boldsymbol{\gamma}^{(m)}, \sigma^{2(m)})\; p(\boldsymbol{\eta}_i \mid \boldsymbol{\Phi}^{(m)}) \qquad (7)$$

Iterating Steps (i) and (ii) is Gibbs sampling from the joint posterior $p(\boldsymbol{\eta}_i, Y_i \mid \boldsymbol{x}_{\text{obs},i}, \boldsymbol{\theta}^{(m)})$, whose $\boldsymbol{\eta}_i$-marginal is the target: the conditioning-model likelihood $p(Y_i \mid \boldsymbol{\eta}_i, \boldsymbol{\gamma}^{(m)}, \sigma^{2(m)})$ is a proper density in $Y_i$, so it integrates to one and drops out of $p(\boldsymbol{\eta}_i \mid \boldsymbol{x}_{\text{obs},i}, \boldsymbol{\theta}^{(m)})$, entering only through $\boldsymbol{\theta}^{(m)}$. The reduced Gibbs sampler iterates $n_{\text{iter}} = 200$ times per observation, reusing the adapted training-stage proposal scale and initialized at the regression factor score (pilot autocorrelation length $\tau$ = 3–8, so burn-in is practically unnecessary); the state reached at the final iteration is taken as that observation's PV under $\boldsymbol{\theta}^{(m)}$, one draw per posterior parameter set rather than an average over iterations. Under the Gaussian measurement model this marginal posterior is analytically Gaussian, so the reduced Gibbs is retained only for extensibility to non-Gaussian models (Section 6.3.3) and is checked against the closed form rather than argued for: `scripts/verify_test_stage_posterior.py` in the archived code compares several hundred independent Step C draws per observation with the analytic mean and covariance, and repeats the comparison under a deliberately wrong $\boldsymbol{\gamma}$ to confirm that the imputed $Y_i$ integrates out. Section 6.2.1 states the practical consequence.

The crux of this procedure is the information separation between training and prediction: $Y$ is used only for parameter estimation in Step B; in Step C, test-set $Y_i$ is unobserved and predictively imputed. No test-set label leakage occurs; training-set $Y$ information enters the test-stage posterior only through $\boldsymbol{\theta}^{(m)}$, the intended mechanism of PV-based predictive inference (cf. Schofield et al. 2015).

## 2.4 Artificial Neural Networks and the Double Ensemble

A shallow feedforward ANN with one hidden layer, ReLU activation, and identity output function is used. The network is optimized with Adam (Kingma and Ba 2015; learning rate = 0.001) on the full training batch, one gradient step per iteration, for at most 500 iterations, with early stopping on a validation-based patience criterion (Prechelt 1998; patience = 10 iterations, with 15% of the training set held out for validation) and the best-validation weights restored at termination. The realized training budget is therefore of the order of several hundred gradient steps; because it is identical for every competing model, it enters all comparisons as a common constant (Section 4.6). The number of hidden nodes and $\lambda_{L2}$ are selected via condition-specific pilot cross-validation (Section 3.5). Study 1 applies L2 to all parameters via PyTorch default `weight_decay`, whereas Study 2 applies L2 only to weight matrices; a sensitivity analysis reported in Section 5.3 shows that this difference does not account for the cross-study pattern difference.

Multiple imputation generates measurement uncertainty through PVs, and the ANN generates algorithmic uncertainty through random weight initialization. To separate these two sources, an $M \times K$ double ensemble is adopted. The final prediction is

$$\hat{Y}_i = \frac{1}{M}\sum_{m=1}^{M}\left(\frac{1}{K}\sum_{k=1}^{K}\hat{f}^{(m,k)}\,(\boldsymbol{\eta}_i^{(m)})\right) \qquad (8)$$

The inner average ($K$) integrates initialization uncertainty, and the outer average ($M$) integrates measurement uncertainty. The variance-decomposition ratio $B/W_K$ is reported by condition (Section 4.1), where $B$ is the between-PV-dataset variance of $K$-averaged predictions and $W_K$ is the within-PV-dataset variance across $K$ initializations, averaged across $M$. The same $K$-ensemble structure is applied to all competing models.

# 3 Study 1: Simulation

## 3.1 Design Overview

The design focuses on core elements—DGP shape, measurement reliability, sample size—to establish the principled advantage of PV-ANN. The population explained variance is fixed at $R^2_{\text{pop}} = .50$, a high-signal condition relative to typical applied settings in personality and educational psychology, where $R^2$ values commonly fall below approximately .10 (Gignac and Szodorai 2016). This choice establishes whether the theoretically predicted advantage materializes under favorable signal conditions before investigating generalization; adding signal strength as a factor would have tripled the number of conditions (from 18 to 54). Generalizability to moderate-signal conditions is examined in Study 2 and discussed in Section 6.3.2. Table 1 summarizes the 18-condition design.

Study 1 was preregistered on the OSF following the ADEMP structure (Morris et al. 2019) prior to executing the simulation; the registration is public at https://osf.io/kfxbw, under the working title "Plausible Values for Nonlinear Prediction Among Latent Variables: A Neural Network Application". Study 2 was not preregistered.

**Deviations from the preregistration.** The data-generating mechanisms, measurement model, sampler settings, and evaluation grid reported below are those preregistered and executed. Six departures are recorded here. (i) *Design.* Sample size was preregistered at the single level $N = 1{,}000$ with $R_{\text{rep}} = 500$; $N = 500$ and $N = 2{,}000$ were added afterwards at $R_{\text{rep}} = 250$ and are reported as an extension rather than as preregistered conditions. (ii) *Added comparators.* Bart-ANN and the two LMS specifications were not in the preregistered method set; they were added to test the score-type invariance of Appendix A.2 and to situate the framework against a parametric benchmark, and Oracle-ANN was recomputed at $K = 15$ rather than the preregistered $K = 3$ to match PV-ANN's ensemble count (Appendix C.7). The preregistration set Sum-ANN, FS-ANN and Oracle-ANN alike at $K = 3$ against PV-ANN's $M \times K = 15$; the equalization of the point-estimate comparators to fifteen networks is reported as a sensitivity analysis in Appendix C.11 rather than applied to the main tables. (iii) *Non-converged replications.* The preregistration specified excluding and replacing replications that failed the convergence gate after five retries; they were instead retained, because exclusion selects on convergence, and Appendix C.4 reports the inclusion-versus-exclusion sensitivity analysis ($|\Delta| \leq .004$) that supports the change. (iv) *Primary metric for RQ 3.* The preregistration made the amplification ratio the decision rule for the amplification hypothesis; because that ratio is unstable as its denominator approaches zero, the difference Diff = NL − Lin is used as the primary metric here, with the ratio reported for all conditions under the preregistered gate (Section 3.6). (v) *Theoretical benchmark.* The preregistration recorded the $k = 2$ amplification benchmark as $1 + 1/\rho$; the value that follows from Equation (A6) is $1 + \rho$, and Appendix A.3 shows that neither transfers directly to the $R^2_{\text{recovery}}$ scale, on which Equation (A6) is read as an ordering prediction only. (vi) *Coverage benchmark.* The preregistration listed 90% coverage of the PV interval among the exploratory diagnostics. It is not reported here, because an interval formed as the range of $M = 5$ draws cannot attain .90: its highest attainable coverage is $(M - 1)/(M + 1) \approx .667$, so the diagnostic cannot discriminate a correctly specified posterior from a misspecified one.

**Predictions formulated after Study 1.** The preregistered recovery hypothesis also predicted lower RMSE for PV-ANN. It did not obtain: PV-ANN RMSE equalled or slightly exceeded FS-ANN in every condition (Section 4.3). Remark 1, which derives that dissociation from the sufficiency of the factor score for predicting $Y$, was formulated in response to this result, so with respect to Study 1 it is an explanation rather than a tested prediction; Study 2, itself not preregistered, provides a check of it on data the account was not built from (Section 5.3). The same status attaches to the positive Diff under the linear DGP, which the preregistration expected to be near zero and which Section 4.5 attributes to two mechanisms operating independently of DGP nonlinearity.

**Table 1** Simulation design

| Factor | Levels |
|---|---|
| DGP | 3 (linear, quadratic, sigmoid) |
| Measurement reliability ($\omega$) | 2 (.94, .80) |
| Sample size ($N$) | 3 (500, 1,000, 2,000) |
| Latent variables | 3 ($\eta_1$: nonlinear main effect + interaction; $\eta_2$: interaction; $\eta_3$: linear control) |
| Number of indicators | 4 per latent variable |
| Number of PVs ($M$) | 5 |
| Initialization ensemble ($K$) | 3 |
| Replications ($R_{\text{rep}}$) | 500 ($N = 1{,}000$) / 250 ($N = 500{,}2{,}000$) |

*Note.* Total conditions = 3 DGP × 2 $\omega$ × 3 $N$ = 18. Training/test split: 80:20. LMS benchmark: Mplus (LMS-Lin and LMS-Q; see Section 3.4.1). At $R_{\text{rep}} = 500$, the Monte Carlo standard error (MCSE; Morris et al. 2019) of $R^2_{\text{recovery}}$ is .0005–.0014 for PV-ANN and .0008–.0026 for FS-ANN, an order of magnitude below the PV-ANN vs. FS-ANN differences of interest (.010–.064). At $R_{\text{rep}} = 250$ the corresponding ranges are .0003–.0044 and .0006–.0064, the maxima occurring in the quadratic $\omega = .80$, $N = 500$ cell.

### 3.2 Data-Generating Process

The structural model is $Y_i = f(\boldsymbol{\eta}_i) + \epsilon_i$:

- **DGP 1 (Linear):** $Y_i = .40\eta_{1i} + .30\eta_{2i} + .20\eta_{3i} + \epsilon_i$
- **DGP 2 (Quadratic):** $Y_i = .30\eta_{1i} + .30\eta_{2i} + .10\eta_{3i} + .15\eta_{1i}^2 + .20\eta_{1i}\eta_{2i} + \epsilon_i$
- **DGP 3 (Sigmoid):** $Y_i = .50\,\text{sigmoid}(5\eta_{1i}) + .30\eta_{2i} + .10\eta_{3i} + .20\,\eta_{2i} \cdot \text{sigmoid}(5\eta_{1i}) + \epsilon_i$

The structural error is $\epsilon_i \sim N(0, \sigma_\epsilon^2)$, with $\sigma_\epsilon^2$ set equal to Var$(f(\boldsymbol{\eta}))$ so that $R^2_{\text{pop}} = .50$ exactly under $\boldsymbol{\eta}_i \sim N(\mathbf{0}, \boldsymbol{\Phi})$, where $\boldsymbol{\Phi}$ specifies a homogeneous latent correlation of .30; the resulting values are $\sigma_\epsilon^2 = .446, .405$, and $.301$ for DGP 1–3. $\Delta R^2_{\text{NL}}$ denotes the increment in explained variance attributable to the nonlinear components: DGP 1 .000, DGP 2 .155, DGP 3 .025 (Appendix C.3).

In DGP 3 the entire $\eta_1$ main effect is carried by the saturating term, so the function is flat in both tails of $\eta_1$ while remaining monotone throughout. This is the property that a symmetric quadratic specification cannot reproduce (Section 4.6; Appendix E), and it makes DGP 3 the complement of DGP 2 in the design: DGP 2 carries a large nonlinear share of explained variance (31%) with a form a quadratic can match, whereas DGP 3 carries a small one (5%) with a form it cannot.

The measurement model specifies four congeneric indicators per latent variable with the standardized loadings shown in Table 2, calibrated to yield $\omega \in \{.94, .80\}$.

**Table 2** Standardized factor loadings and composite reliability

| Condition | $\boldsymbol{\lambda_1^*}$ | $\boldsymbol{\lambda_2^*}$ | $\boldsymbol{\lambda_3^*}$ | $\boldsymbol{\lambda_4^*}$ | $\boldsymbol{\omega}$ | $\boldsymbol{\rho}$ |
|---|---|---|---|---|---|---|
| High reliability | .85[a] | .88 | .91 | .94 | .94 | .949 |
| Low reliability | .60[a] | .67 | .74 | .81 | .80 | .818 |

*Note.* Based on the standardized solution. $\psi_j^* = 1 - \lambda_j^{*2}$. $\omega$ = composite reliability (McDonald 1999). $\rho$ = factor score coefficient of determination (Equation (2)). [a]Marker variable.

**3.3 Gibbs Sampler**

The Gibbs sampler implements four blocks following the general Bayesian structural equation modeling framework (Lee 2007):

1. $\boldsymbol{\Lambda}, \boldsymbol{\Psi} \mid \boldsymbol{\eta}, \boldsymbol{X}$: Indicator-wise normal-inverse-gamma (NIG). Marker-variable loading fixed. Hyperparameters: $\alpha_0 = 1$, $\beta_0 = 1$, $V_0 = 10$.
2. $\boldsymbol{\gamma}, \sigma^2 \mid \boldsymbol{\eta}, \boldsymbol{Y}$: Normal-inverse-gamma. $\boldsymbol{\gamma} \mid \sigma^2 \sim N(\mathbf{0}, \sigma^2 \cdot 100\boldsymbol{I})$, $\sigma^2 \sim$ IG(0.01,0.01).
3. $\eta_i \mid \cdot$: Adaptive Metropolis-Hastings (MH). The proposal covariance is $(\boldsymbol{\Lambda}'\boldsymbol{\Psi}^{-1}\boldsymbol{\Lambda} + \boldsymbol{\Phi}^{-1})^{-1} \cdot s^2$. During the first 80% of burn-in, $\log s$ is updated via Robbins-Monro (Robbins and Monro 1951) targeting an acceptance rate of .234 (Roberts et al. 1997), with adaptation fixed during the final 20%. The observed median acceptance rate of .270–.278 falls between this asymptotic target and the finite-$P$ optimum (~.40; Roberts and Rosenthal 2001); pilot experiments with a target of .40 yielded a comparable effective sample size (ESS).
4. $\boldsymbol{\Phi} \mid \boldsymbol{\eta}$: Inverse-Wishart. $\nu_0 = P + 2 = 5$, $\boldsymbol{V}_0 = \boldsymbol{I}_P$.

Chain length is 3,000 burn-in and 15,000 post-burn-in iterations in Study 1 (extended in Study 2; Section 5.2). $M = 5$ PVs are drawn at equal intervals starting at the 5% point of the post-burn-in window. Convergence is assessed on the conditioning-model chains ($\boldsymbol{\gamma}$, $\sigma^2$) from a single chain split into two halves, with criteria split-$\hat{R} < 1.05$ in its classical, non-rank-normalized form (Gelman and Rubin 1992) and ESS $> 200$ from the initial-positive-sequence estimator (Vehtari et al. 2021).

## 3.4 Competing Models

Table 3 lists the prediction models compared in the simulation.

**Table 3** Prediction models compared in the simulation

| | **OLS** | **QR (quadratic)** | **ANN (× $K$)** |
|---|---|---|---|
| **Sum score** | Sum-LR (1) | — | Sum-ANN (3) |
| **Regression factor score** | FS-LR (1) | — | FS-ANN (3) |
| **Bartlett factor score** | — | — | Bart-ANN (3) |
| **PV** | PV-LR (5) | PV-QR (5) | **PV-ANN (15)** |
| **True $\eta$** | — | — | Oracle-ANN (15) |

*Note.* OLS = ordinary least squares. QR = quadratic regression. Model names combine an input prefix (Sum = sum score; FS = regression factor score; Bart = Bartlett factor score; PV = plausible values; Oracle = true $\boldsymbol{\eta}$) with an estimator suffix (LR = linear regression, estimated by OLS; QR; ANN). Numbers in parentheses indicate the total number of prediction models trained per replication (e.g., PV-ANN trains $M \times K = 5 \times 3 = 15$ models). Oracle-ANN is reported on a $K = 15$ basis, matching PV-ANN's total ensemble count; the main run used $K = 3$ and all Oracle results were recomputed at $K = 15$ (Appendix C.7). Bart-ANN and FS-ANN alike use population-parameter-based scores and are therefore free of parameter estimation error (Section 3.4); the attainable upper bound for point-estimate-based learners is the population-optimal ceiling of Appendix C.3.

**Input.** All inputs for ANN- and OLS-based models are $z$-scored based on the training set before entry. Regression factor scores and Bartlett factor scores were computed using population parameters; because these scores carry no parameter estimation error, the margin by which PV-ANN exceeds them is a lower bound on the advantage attributable to integrating measurement uncertainty. ANN hyperparameters (hereafter HP) are selected on PV-ANN's hold-out RMSE and applied identically to all competing models (Section 3.5). These two choices bias the comparison in opposite directions—parameter-free scores favor the competitors, shared HPs may favor PV-ANN—so the net direction is not fixed a priori; the Bart-ANN benchmark bears on this point (Section 6.1).

### *3.4.1 LMS Benchmark*

LMS (Klein and Moosbrugger 2000) was implemented using `TYPE = RANDOM` in Mplus 9 (Muthén and Muthén 2023). Two specifications were applied: LMS-Lin (`Y ON eta1 eta2 eta3`), a linear baseline; and LMS-Q (`Y ON eta1 eta2 eta3 int12 quad1`), adding the $\eta_1 \times \eta_2$ interaction and the $\eta_1^2$ term. LMS-Q functions as overspecification under DGP 1, correct specification under DGP 2, and misspecification under DGP 3. A structural asymmetry exists in the $R^2_{\text{recovery}}$ computation between LMS and PV-ANN, favoring LMS under correct specification (DGP 2) and neutralized under misspecification (DGP 3); LMS RMSE further reflects a scale mismatch from applying latent-scale coefficients to attenuated factor-score inputs. Both points are detailed in Appendix E.

Following Section 3.6, unstandardized structural coefficients from Mplus are converted to the standardized scale using population latent variances before grid substitution. RMSE computation uses regression factor scores extracted from the test set as the input to the LMS-estimated function. Mplus convergence settings and rates are reported in Table D1.

### 3.5 Preliminary Tuning and Convergence Quality Control

MCMC pilots identified the conditioning model coefficient $\gamma_2$ as the convergence bottleneck in the worst-case condition (DGP 3, $\omega = .80$), with $\tau \approx 52–90$, attributable to the dependence structure between the $\boldsymbol{\gamma}$–$\boldsymbol{\eta}$ blocks. A $\hat{R}$/ESS quality-control gate was applied: chains failing the criteria were rerun with a new seed (up to 5 attempts); the final attempt's result was used if criteria remained unmet, and the frequency is reported in Section 4.1.

ANN HPs were selected from six combinations of hidden nodes $\in \{8,16\} \times \lambda_{L2} \in \{10^{-4}, 10^{-3}, 10^{-2}\}$ over 20 pilot replications per condition, minimizing PV-ANN's hold-out RMSE (Table C1). Recomputing the main-run summaries without those 20 replications moves RMSE by at most .0012 and $R^2_{\text{recovery}}$ by at most .0016 in any of the 18 conditions, so the pilot–main-run overlap does not carry a practically meaningful optimistic bias; the criterion mismatch between RMSE selection and $R^2_{\text{recovery}}$ evaluation works in the same direction (Appendix C.5, Table C3). $K$ was fixed at 3: RMSE decreased considerably from $K = 1$ to 3, while the additional decrease from $K = 3$ to 5 was marginal. Sufficiency of $M = 5$ is supported by the small |ΔRMSE| and $|\Delta R^2|$ between $M = 5$ and $M = 3$ (Section 4.1). Fraction of missing information (FMI) is reported descriptively only, as Rubin's (1987) MI efficiency formula was derived for linear complete-data analyses.

### 3.6 Evaluation Metrics

$R^2_{\text{recovery}}$ **(unweighted).** Goodness of fit on the true $\eta$ grid:

$$R^2_{\text{recovery}} = 1 - \frac{\sum_g [f(\boldsymbol{\eta}_g) - \hat{f}(\boldsymbol{\eta}_g)]^2}{\sum_g [f(\boldsymbol{\eta}_g) - \bar{f}]^2} \qquad (9)$$

where $\bar{f} = \frac{1}{G}\sum_g f(\boldsymbol{\eta}_g)$ is the mean of true function values at the $G = 729$ grid points ($9^3$ equally spaced points in $[-2, +2]^3$).

The estimand is each model's implied regression function on the latent scale, $E[\hat{Y} \mid \boldsymbol{\eta}]$, compared with $f$. The question is what a model predicts for a case whose true latent value is $\boldsymbol{\eta}_g$—not what it predicts at a given value of its own input—and that fixes the mapping rather than leaving it to be chosen. For PV-ANN, $\hat{f}(\boldsymbol{\eta}_g) = \frac{1}{MK}\sum_{m,k} \hat{f}^{(m,k)}(\boldsymbol{\eta}_g)$. For a score-based model, a grid point $\boldsymbol{\eta}_g$ is mapped to the score such a case would be expected to obtain, $E[\boldsymbol{s} \mid \boldsymbol{\eta} = \boldsymbol{\eta}_g]$, and then standardized with the model's training scaler (Section 3.4). For a linear score with coefficient of determination $\rho_s$ this places the model at the standardized coordinate $\sqrt{\rho_s}\,\boldsymbol{\eta}_g$; regression and Bartlett scores therefore land on exactly the same coordinate, the sum score on a slightly contracted one, and PV- and Oracle-based models ($\rho_s = 1$) on $\boldsymbol{\eta}_g$ itself. Evaluating every model at a common *standardized* coordinate instead measures a different quantity, in which the $k$th-order coefficient is attenuated by $\rho_s^{k/2}$ rather than $\rho_s^k$; Appendix A.3 sets out the two conventions and shows that the choice rescales the gaps without reordering the models.

$R^2_{\text{recovery},w}$ **(density-weighted).** The multivariate standard normal density $\phi(\boldsymbol{\eta}_g)$ is applied as weights:

$$R^2_{\text{recovery},w} = 1 - \frac{\sum_g \phi(\boldsymbol{\eta}_g)[f(\boldsymbol{\eta}_g) - \hat{f}(\boldsymbol{\eta}_g)]^2}{\sum_g \phi(\boldsymbol{\eta}_g)[f(\boldsymbol{\eta}_g) - \bar{f}_w]^2} \qquad (10)$$

where $\bar{f}_w$ is the density-weighted mean.

**RMSE.** Hold-out test set prediction error. Because comparisons span numerous conditions, interpretation is based on the consistency of patterns (Skrondal 2000), with 95% bootstrap CIs as supplementary. Under Remark 1, RMSE additionally serves as a falsification check: the theory predicts parity or a small PV deficit on this metric, so an observed PV advantage would signal a mechanism outside the theory.

**Amplification difference.** The metric for RQ 3 is Diff = (PV-ANN − Sum-ANN) − (PV-LR − Sum-LR) in $R^2_{\text{recovery}}$, computed within each (DGP, $\omega$, $N$) cell; equivalently, Diff = NL − Lin. Under the linear DGP, Diff captures predictor-type amplification independent of DGP nonlinearity and serves as a baseline. Two baselines are reported: the sum score, the default-practice comparator, and the regression factor score, the theory-matched one, since the $\rho^k$ bound of Section 1.1 is stated in terms of its coefficient of determination; they support the same conclusions (Table D5 panels (a) and (b)). Amp = NL / Lin, the preregistered ratio form, is reported in Table D5 for completeness and is not interpreted (Section 3.1(iv)).

### 3.7 Computing Environment

Python 3.11, NumPy 1.26, PyTorch 2.5 (Paszke et al. 2019; CPU, deterministic algorithm mode), and pandas 2.0. LMS was run with Mplus 9 (Muthén and Muthén 2023). End-to-end wall time on a six-worker CPU setup was approximately 101 h for the Study 1 main run and 6.1 h for the Study 2 main analysis, plus 42.6 h for the Study 2 LMS benchmark and 12.0 h for the regularization sensitivity run of Section 5.3. Per-replication wall times are archived, and the reported totals are their sums. Code, seeds, pilot diagnostics, and a README documenting execution order are deposited with the analysis package (Data and Code Availability).

## 4 Study 1: Results

### 4.1 Convergence Diagnostics

In the nine $\omega = .94$ conditions, the Gibbs sampler converged in every replication (rerun rate = 0%). The $\omega = .80$ conditions failed the gate frequently: rerun rates were 30.0%–75.2%, with a failure rate of 44.4% in the worst case (sigmoid, $\omega = .80$, $N = 500$). Three facts determine how those failures should be read. No failed replication showed $\hat{R}_{\max} > 1.10$; failures were driven predominantly by the effective-sample-size criterion (failed-replication ESS medians 124–162 across the $\omega = .80$ conditions, archived with the deposited diagnostics; the corresponding all-replication medians are 203–254, Table C6) rather than by $\hat{R}$; and in the worst diagnostic condition at $N = 1{,}000$ (sigmoid, $\omega = .80$), doubling the post-burn-in length (15,000 → 30,000) eliminated failures entirely (23.0% → 0.0%) while leaving RMSE and $R^2_{\text{recovery}}$ unchanged within MCSE (Table C4). The default chain was therefore too short rather than the sampler pathological, the bottleneck being $\gamma_2$ (Section 3.5). Inclusion vs. exclusion of unconverged replications changed results by $|\Delta| \leq .004$, far smaller than the PV-ANN advantage over FS-ANN ($\geq .011$ across $\omega = .80$ conditions), indicating that the advantage direction is robust to convergence status (Tables C2, C6). The extended chain was not adopted for the main run because doubling the post-burn-in length doubles the dominant computational cost across all 18 conditions; the sensitivity analyses above bound what that choice costs. LMS-Lin and LMS-Q achieved 100% convergence across all conditions (Table D1).

Condition-specific HPs, FMI, and variance decomposition are reported in Table C1. $B/W_K$ exceeded 1 in all conditions and was 11.8 in the quadratic $\omega = .80$, $N = 1{,}000$ condition (maximum 18.0; Table C1), indicating that measurement uncertainty dominated algorithmic uncertainty in low-reliability nonlinear conditions. The $M = 5$ versus $M = 3$ differences were $|\Delta\text{RMSE}| \leq .007$ and $|\Delta R^2| \leq .006$, supporting the sufficiency of $M = 5$.

### 4.2 Posterior Diagnostics: Empirical Evaluation of Congeniality

At the population level, the proportion of $f(\boldsymbol{\eta})$ variance explained by the second-order polynomial conditioning model is 100% for DGP 2 and 96.8% for DGP 3 (Appendix C.3). The unexplained 3.2% in DGP 3 is attributable to higher-order components of the sigmoid function and may produce locally elevated PV bias in the transition region ($|\eta_1| < 1$).

PV bias by true $\eta_1$ interval is approximately monotone in every condition, positive below the mean and negative above it, and its magnitude at $\omega = .80$ is approximately 3–4 times that at $\omega = .94$ (Table C7). This is the signature of Bayesian posterior shrinkage under finite measurement information, which the present design cannot separate from congeniality-induced distortion. The separation is not needed: PV-ANN's recovery ratio against the Oracle, which sees the true $\boldsymbol{\eta}$ and is free of PV-induced distortion, remained near unity in every condition (Section 4.3), which bounds the net impact of both sources on $R^2_{\text{recovery}}$.

### 4.3 RQ 1: Do PVs Improve Nonlinear Recovery?

In nonlinear DGPs, PV-ANN consistently improved $R^2_{\text{recovery}}$ relative to point-estimate-based ANNs (Table 4). The recovery ratio relative to the Oracle ($K = 15$) was .972–1.001, supporting the theoretical prediction (Section 1.1) that PV variance preservation substantially eliminates $\rho^k$ attenuation. The Oracle is subject to the same training budget as every other model (Section 2.4), so this ratio measures how closely PV-ANN approaches a learner that receives the true $\boldsymbol{\eta}$ under matched estimation conditions, not how closely it approaches the population-optimal function.

**Table 4** $R^2_{\text{recovery}}$: $N = 1{,}000$ conditions
**(a) Run of record** ($R_{\text{rep}} = 500$)

| DGP | $\boldsymbol{\omega}$ | Oracle | PV-ANN | FS-ANN | Sum-ANN | Bart-ANN | PV-QR | LMS-Lin | LMS-Q |
|---|---|---|---|---|---|---|---|---|---|
| Linear | .94 | .9822 | .9812 | .9677 | .9678 | .9686 | .9626 | .9896 | .9824 |
| Linear | .80 | .9832 | .9715 | .9514 | .9472 | .9514 | .9245 | .9850 | .9750 |
| Quadratic | .94 | .9424 | .9411 | .9224 | .9203 | .9245 | .9632 | .6490 | .9826 |
| Quadratic | .80 | .9421 | .9320 | .8677 | .8616 | .8726 | .9232 | .6454 | .9718 |
| Sigmoid | .94 | .9507 | .9478 | .9376 | .9370 | .9382 | .9317 | .9268 | .9539 |
| Sigmoid | .80 | .9450 | .9373 | .8994 | .8960 | .9023 | .8962 | .9215 | .9453 |

**(b) Point-estimate learners equalized to $K = 15$**

| DGP | $\omega$ | FS-ANN | Sum-ANN | Bart-ANN | PV − FS | PV − Sum | PV − Bart |
|---|---|---|---|---|---|---|---|
| Linear | .94 | .9773 | .9774 | .9792 | .0039 | .0038 | .0020 |
| Linear | .80 | .9591 | .9540 | .9579 | .0124 | .0175 | .0136 |
| Quadratic | .94 | .9288 | .9257 | .9289 | .0123 | .0154 | .0122 |
| Quadratic | .80 | .8721 | .8695 | .8776 | .0599 | .0625 | .0544 |
| Sigmoid | .94 | .9448 | .9453 | .9452 | .0030 | .0025 | .0026 |
| Sigmoid | .80 | .9073 | .9016 | .9088 | .0300 | .0357 | .0285 |

*Note.* Oracle = Oracle-ANN ($K = 15$). Linear-predictor results (PV-LR, FS-LR, Sum-LR) and all 18 conditions are reported in Table D2; panel (b) for all 18 is in Table C9. MCSE in panel (a) ranges .0003–.0028. PV-ANN averages $M \times K = 15$ networks, so panel (a) compares it against the $K = 3$ ensembles of Table 3 and panel (b) against fifteen networks each. Panel (b) is formed as the panel (a) value plus the within-replication $K = 15$ minus $K = 3$ difference of Appendix C.11, whose paired $SE$ (.0009–.0025 in these cells) is an order of magnitude below that of either arm's mean. Both panels are reported throughout: panel (a) is the preregistered run, panel (b) removes the one design asymmetry that would otherwise qualify every gap.

In the $\omega = .80$ nonlinear conditions, the PV-ANN advantage was +.064 for the quadratic and +.038 for the sigmoid DGP against FS-ANN, and +.060 and +.030 with the ensembles equalized. Read against the Oracle rather than in absolute units, PV-ANN closes 86% (quadratic) and 83% (sigmoid) of the distance separating FS-ANN from a learner that receives the true $\boldsymbol{\eta}$, and 77–91% of it across the $N \geq 1{,}000$ nonlinear $\omega = .80$ conditions (`scripts/export_recovery_share.py` in the archived code). The share is the more stable form of the result because numerator and denominator absorb the ensemble convention together: equalization moves it by 0.6–4.3% of its own value against 6.4–20.8% for the absolute gap. Bart-ANN was comparable to or slightly superior to FS-ANN yet substantively inferior to PV-ANN on either convention, consistent with the score-type invariance of the recovery limit (Appendix A.2). Sum-ANN performed slightly below FS-ANN in most Study 1 conditions (differences within .009), reflecting loss of indicator-specific optimal weighting under heterogeneous loadings—a Study 1 ordering that contrasts with Sum-ANN's edge in Study 2 (Section 5.3).

**Ceiling check: the size of the gaps, not only their direction.** Evaluating the population-optimal score-based predictor $E[f(\boldsymbol{\eta}) \mid \boldsymbol{s}]$ on the same grid gives the ceiling attainable by any point-estimate-based learner, and $1 -$ ceiling is therefore the PV advantage that $\rho^k$ attenuation alone implies. Because the ceilings are population quantities they do not depend on $N$; the comparison is made at $N = 2{,}000$, where finite-sample estimation error is smallest (Table 5). The observed gaps track the implied ones to within $.010$ as reported and $.007$ equalized, in every DGP, and the ceilings for regression and Bartlett scores agree to four decimal places in every condition, which is the numerical counterpart of Appendix A.2. This is the quantitative form of the test: a learner limited by something other than $\rho^k$ attenuation—an undertrained network, say, or a regularization artifact—would not reproduce three DGP-specific magnitudes that the attenuation law fixes in advance.

**Table 5** Observed PV advantage against the ceiling implied by $\rho^k$ attenuation ($\omega = .80$, $N = 2{,}000$)

| DGP | Score ceiling | Implied PV advantage | Observed | Observed, equalized |
|---|---|---|---|---|
| Linear | .982 | .018 | .028 | .022 |
| Quadratic | .947 | .053 | .059 | .056 |
| Sigmoid | .967 | .033 | .030 | .026 |

*Note.* Ceiling = $R^2_{\text{recovery}}$ of $E[f(\boldsymbol{\eta}) \mid \boldsymbol{s}]$ for the regression factor score, evaluated on the grid under the rule of Section 3.6; the Oracle's ceiling is 1.000, so the implied PV advantage is its complement (Appendix C.3). Observed = PV-ANN − FS-ANN from Table D2; equalized as in Table 4 panel (b). Values are produced by `python scripts/export_population_ceiling.py` in the archived code. The linear-DGP excess of $.010$ falls to $.004$ under equalization, which is the part Section 4.5 attributes to mechanisms independent of $\rho^k$ attenuation.

**LMS benchmark.** In DGP 2 (correct specification), LMS-Q exceeded PV-ANN by .042 / .040 ($\omega = .94$ / $\omega = .80$), reflecting the parametric efficiency of correct specification (structural asymmetry detailed in Appendix E). In DGP 3 (misspecification, $N = 1{,}000$), the LMS-Q advantage was reduced to .006 / .008—80–86% reductions—and became nonsignificant or reversed in some condition–metric combinations (Tables D2 and D3); both estimators fail to fully capture the sigmoid saturation (Section 4.6). In DGP 1 (overspecification), adding the quadratic term in LMS-Q produced negative differences of $-.003$ to $-.022$ across all conditions.

**Directional divergence between $R^2_{\text{recovery}}$ and RMSE.** PV-ANN RMSE was mostly slightly higher than FS-ANN ($|\Delta| \leq .007$; Table D4). A bias–variance decomposition (Appendix G) shows that PVs substantially reduce Bias$^2$ by removing $\rho^k$ attenuation but introduce additional between-imputation variance ($V_{\text{between}}$) that offsets this benefit on RMSE. This dissociation—a PV advantage on $R^2_{\text{recovery}}$ alongside parity to a slight deficit on RMSE—is exactly the pattern predicted by Remark 1, supporting prediction (ii).

### 4.4 RQ 2: Does the Advantage Increase With Greater Measurement Error?

RQ 2 was supported for function shape recovery. At $N = 1{,}000$ the $R^2_{\text{recovery}}$ advantage of PV-ANN over FS-ANN was .064 (quadratic) and .038 (sigmoid) at $\omega = .80$, against .019 and .010 at $\omega = .94$—an increase across reliability levels of .046 and .028. Equalized, the same four gaps are .060, .030, .012 and .003 (Table 4 panel (b)), so the increase is essentially unchanged (.048 and .027) on a high-reliability baseline that has dropped close to zero, which is what the $\rho^k$ arithmetic of Section 1.1 predicts. The scaling is stated as this difference rather than as the ratio of the two gaps, for the reason Section 3.1 gives for the amplification ratio: the equalized $\omega = .94$ gap for the sigmoid DGP is .003 with a paired standard error of .0014 (Table C9), so a ratio formed on it is not estimable. The same pattern was observed for $R^2_{\text{recovery},w}$ (Table D3). For RMSE, the difference between PV-ANN and FS-ANN was small ($|\Delta| \leq .007$).

The advantage also varies with sample size, though not monotonically: $R^2_{\text{recovery}}$ rises with $N$ for every model (Table D2), while in the nonlinear conditions the PV-ANN margin over FS-ANN peaks near $N = 1{,}000$ at $\omega = .80$ and declines monotonically with $N$ at $\omega = .94$, where high reliability leaves little for PVs to restore. The $N = 500$, $\omega = .80$ conditions are read as exploratory reference values because of the elevated convergence failure rate there (Section 4.1).

### 4.5 RQ 3: Is the PV Advantage Amplified in Nonlinear Prediction?

The amplification difference (Diff = NL − Lin) was positive in all 18 conditions with 95% CIs excluding zero, on both the sum and the FS baseline (Table D5). At $\omega = .80$, $N = 1{,}000$ it was +.062 (quadratic) and +.031 (sigmoid) against +.013 under the linear DGP, and all three remain positive with the ensembles equalized.

That linear-DGP value is a baseline rather than an effect: there is no nonlinear component there to amplify. Two mechanisms independent of DGP nonlinearity produce it. The larger is ensemble size—equalizing it moves the six $N = 1{,}000$ Diff values from +.012–.062 to +.004–.054, the linear-DGP cell losing most of its own value (Table 4 panel (b); Appendix C.11). The remainder is that inclusive conditioning on $Y_{\text{train}}$ (Section 2.2) carries outcome-relevant information into the posterior draws that population-parameter factor scores cannot reflect, which reaches the linear predictors as well: PV-LR attains $R^2_{\text{recovery}} = .982$ against FS-LR's $.975$ in the linear $\omega = .80$, $N = 1{,}000$ cell (Table D2). Subtracting the linear-DGP baseline is what isolates the nonlinear contrast.

RQ 3 is accordingly read as a consistency check on the $\rho^k$ law rather than as an independent test of it. What the design supports is the $k \geq 2$ versus $k = 1$ contrast of Equation (A6), which RQ 1 and RQ 2 establish more directly and with less machinery. The difference between the two nonlinear DGPs is not evidence on the order of the component, since they differ also in its share of explained variance ($\Delta R^2_{\text{NL}}$: .155 vs. .025). The numerical value of $1 - \rho^k$ is not a target on the $R^2_{\text{recovery}}$ scale (Appendix A.3). And the preregistered ratio form Amp = NL / Lin, reported for completeness in Table D5, is interpretable in only 6 of the 18 conditions and is not interpreted here (Section 3.1(iv)).

**Role of PV-QR.** PV-QR outperformed PV-ANN in DGP 2, $\omega = .94$ (.963 vs. .941) but fell short in small-sample $\omega = .80$ conditions. In DGP 3, PV-ANN exceeded PV-QR in all conditions, indicating that ANN's form-free learning provides an advantage when the form is unknown.

### 4.6 Function Shape

Fig. 1 displays the contour recovery for DGP 2 ($\omega = .80$, $N = 1{,}000$). PV-ANN and LMS-Q both recover the curved surface structure of the true DGP; LMS-Q approaches the true extrema more closely, reflecting correct parametric specification. PV-ANN exhibits modest attenuation at the extrema, though substantially less than FS-ANN, whose compressed value range visually reflects the variance-shrinkage attenuation.

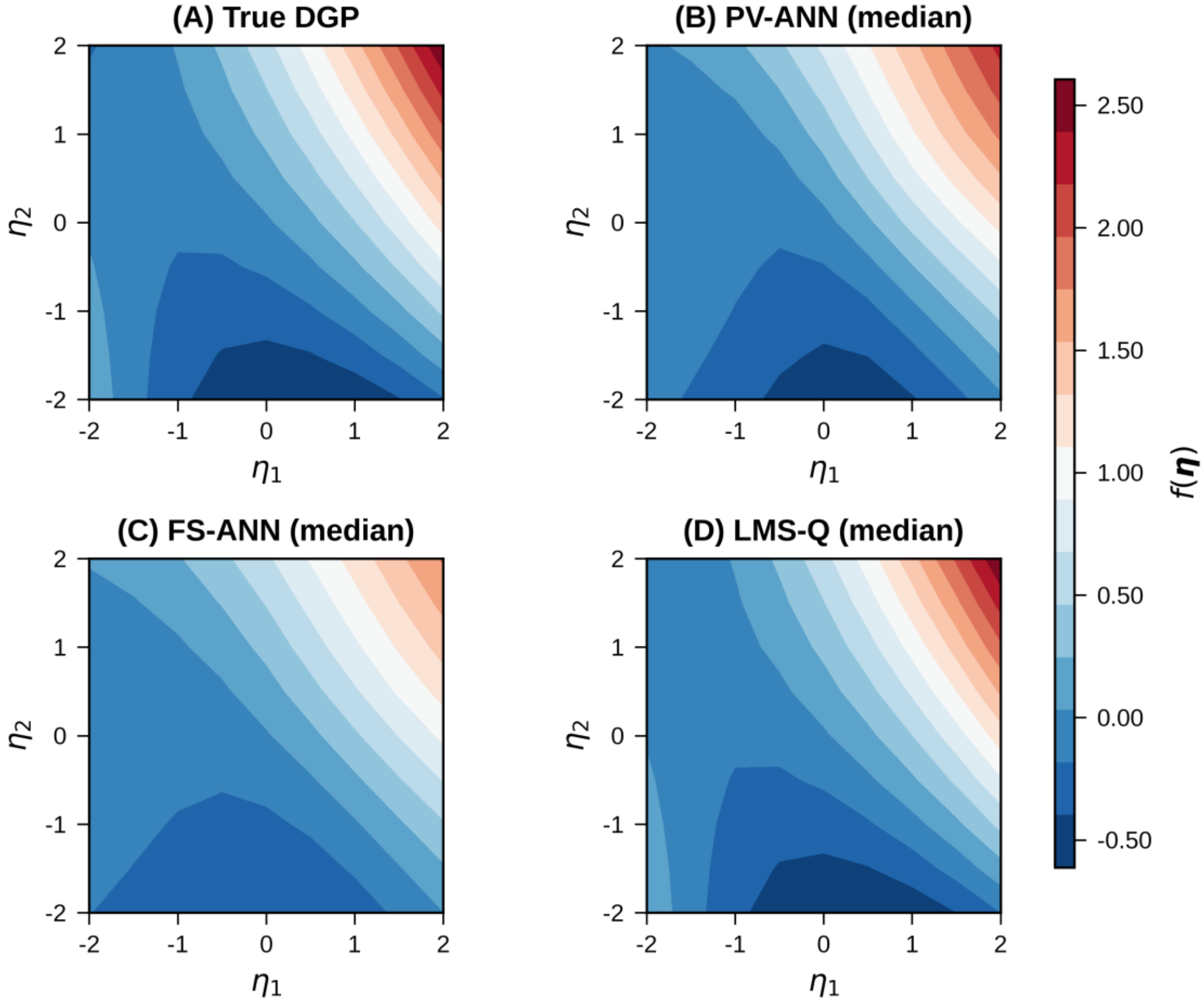

**Fig. 1** True surface and model-specific recovery contour plot for the DGP 2, $\omega = .80$, $N = 1{,}000$ condition. Contours show $f(\boldsymbol{\eta})$ over $(\eta_1, \eta_2)$ with $\eta_3 = 0$ held fixed. Median of 500 replications. Panels: (A) True DGP, (B) PV-ANN, (C) FS-ANN, (D) LMS-Q. FS-ANN illustrates the variance-shrinkage attenuation of point-estimate-based ANNs; LMS-Q provides the parametric benchmark under correct specification. A supplementary figure displaying Oracle-ANN, Sum-ANN, Bart-ANN, PV-QR, and LMS-Lin is provided in Fig. D1

Fig. 2 shows the $\eta_1$ slice recovery for DGP 3 ($\omega = .80$, $N = 1{,}000$). All ANN-based estimators—including Oracle-ANN—and LMS-Q fail to fully capture the sigmoid saturation in the tails, with comparable residuals across estimators in the transition region ($|\eta_1| < 1$). The parallel failure of Oracle-ANN, which sees the true $\boldsymbol{\eta}$, shows that this gap is not measurement-error-induced attenuation but an approximation floor of the shared estimation procedure. Supplementary runs varying the hidden width, the L2 penalty, and the early-stopping budget of Section 2.4, singly and jointly, reduce the residual by at most about a third, none removes it, and all move the residual in the tails and at the center of the grid in nearly the same proportion (Appendix C.10). Because the estimation setting is held constant across models by design, the floor shifts all estimators together and leaves their ordering unaffected. The PV-ANN advantage over LMS-Lin and FS-ANN accordingly manifests in the density-weighted metric (Table D3) rather than in a visually distinct residual structure.

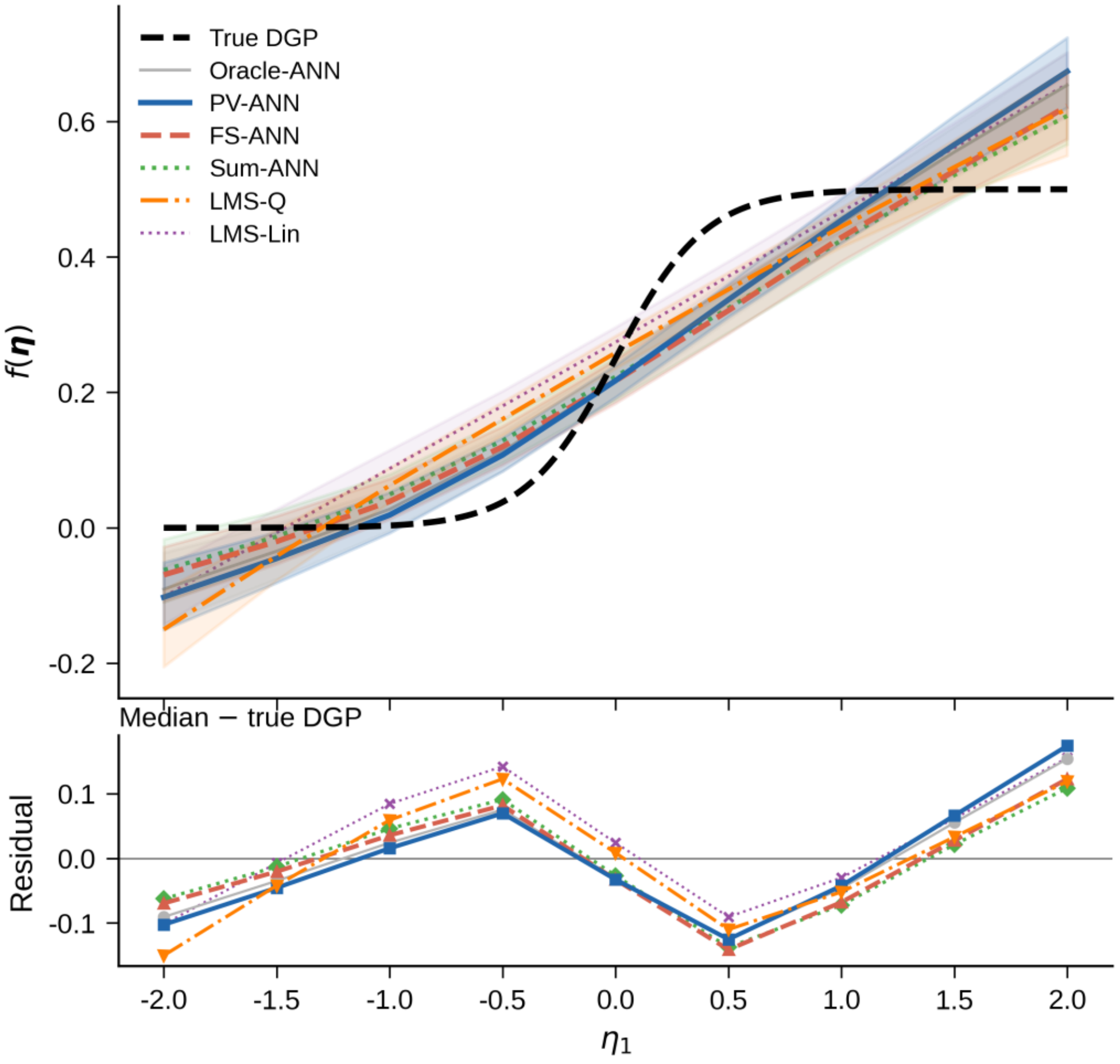

**Fig. 2** $\eta_1$ slice recovery curves for the DGP 3, $\omega = .80$, $N = 1{,}000$ condition. $\eta_2 = \eta_3 = 0$ fixed. Oracle-ANN is trained with $K = 15$ independent initializations. Shaded regions in the upper panel represent pointwise 2.5/97.5 percentile bands across 500 replications. The lower panel displays the residual (median prediction − true DGP) at each $\eta_1$ grid point

# 5 Study 2: Empirical Application

## 5.1 Data

A random subsample of $N = 10{,}000$ was drawn from the International Personality Item Pool (IPIP) Big Five dataset (Goldberg 1999; Goldberg et al. 2006) at the Open-Source Psychometrics Project (https://openpsychometrics.org/_rawdata/). The independent variables (IVs) are three factors—Extraversion ($\eta_1$), Emotional Stability ($\eta_2$), and Agreeableness ($\eta_3$)—each measured by 10 items; the dependent variable (DV) is Openness (10-item sum score, treated as continuous). The DV enters as a sum score, mirroring the simulation design in which $Y$ carries no separate measurement model; its measurement error is absorbed into the structural error and does not affect the present focus on IV-side error correction. This absorption deflates $R^2_{\text{total}}$. In the analytic sample the Openness scale's composite reliability is $\omega_Y = .802$ (Cronbach's $\alpha = .802$; 10 items, congeneric maximum-likelihood CFA fitted as for the independent variables), so a DV-side-only attenuation correction gives $R^2_{\text{latent}} \approx R^2_{\text{obs}}/\omega_Y \approx .067$, a lower bound since IV-side attenuation further deflates the observed $R^2$. Five-point Likert items were treated as continuous, consistent with Rhemtulla et al. (2012); extension to an ordinal measurement model is discussed in Section 6.3.3. Descriptive statistics are reported in Table B1.

## 5.2 Measurement Model

**Table 6** CFA results and reliability

| Factor | No. items | $\lambda^*$ range | $\omega$ | $\rho$ |
|---|---|---|---|---|
| Extraversion ($\eta_1$) | 10 | [.585, .766] | .902 | .909 |
| Emot. Stability ($\eta_2$) | 10 | [.433, .760] | .877 | .889 |
| Agreeableness ($\eta_3$) | 10 | [.362, .775] | .849 | .876 |

*Note.* $N = 10{,}000$. CFA fit: CFI = .815, TLI = .800, RMSEA = .078, SRMR = .084. Latent correlation range: .01–.37. The $\omega$ range (.849–.902) falls between the two simulation conditions (.94, .80). The ρ column reports the diagonal of the multivariate coefficient-of-determination matrix (Equation (C1)); the corresponding univariate values (Equation (2)) are .908, .888, and .874.

A congeneric CFA model was specified with three correlated factors, each measured by 10 items, with no correlated residuals or cross-loadings. CFA fit (Table 6) fell below conventional cutoffs (Hu and Bentler 1999), indicating substantive misspecification; the usual sources are cross-loadings, correlated residuals among similarly worded items, and reverse-scoring method factors, and the number-of-indicators effect on CFI (Kenny and McCoach 2003) may account for part of the shortfall, values in the .80–.85 range being common for 10-item personality scales under strict congeneric assumptions (Hopwood and Donnellan 2010). Whatever the source, misspecification degrades PV posterior quality because PVs are drawn conditional on the assumed model; implications are discussed in Section 6.3.2. In Study 2, regression and Bartlett factor scores entering FS-ANN and Bart-ANN were computed from maximum-likelihood CFA parameter estimates obtained once on the full analytic sample ($N = 10{,}000$), since population parameters are unavailable in field data; the same estimates supplied the Gibbs sampler's marker-variable parameterization and initial values. Because the CFA uses item responses only, no outcome information crosses holdout boundaries. This differs from Study 1, where population-parameter scores served as an upper bound for point-estimate-based approaches (Section 3.4). To accommodate the larger measurement model (10 items per factor), the Gibbs chain was extended relative to Study 1 to 5,000 burn-in and 25,000 post-burn-in iterations, with all other sampler settings, priors, and convergence criteria identical (Section 3.3).

## 5.3 Results

ANN hyperparameters were selected via pilot cross-validation on the first 20 of the 30 holdout splits—mirroring Study 1's protocol and its pilot–main-run overlap structure (Appendix C.5)—with the hidden-node grid expanded to $\{4,8,16\}$ for the weaker-signal setting (cf. Section 6.2.1); the selected configuration was hidden nodes = 16, $\lambda_{L2} = .001$. For ANN optimization only, $Y$ was $z$-standardized on each training split and predictions were back-transformed, because the raw Openness scale ($M \approx 39$) impedes Adam convergence at the common learning rate; closed-form linear models used raw $Y$, and Study 1's unit-scale $Y$ required no rescaling. The evaluation logic follows RQ 4 and Remark 1. Because true latent function values are unobservable in field data, $R^2_{\text{recovery}}$ cannot be computed; the shape of the estimated function is instead examined through the partial dependence plot (PDP) in Fig. 3, which carries the qualitative evidence in Study 2. The PDP is a shape diagnostic, not a field measurement of $R^2_{\text{recovery}}$: its axis is the standardized score, the first of the two conventions of Appendix A.3, and Appendix H.3 re-expresses the archived curves on the Section 3.6 coordinate without retraining. Hold-out RMSE and cross-validated $R^2$ (CV $R^2$; 30 repeated holdouts) correspond to Study 1's secondary metric, on which Remark 1 predicts parity among PV- and point-estimate-based learners; the CV $R^2$ comparison thus functions as a specification check of the theory rather than as the primary comparison.

**Table 7** Hold-out RMSE and cross-validated $R^2$

| **Model** | **RMSE $M$ ($SD$)** | **CV $R^2$ $M$ ($SD$)** |
|---|---|---|
| PV-ANN | 6.019 (.085) | .054 (.008) |
| FS-ANN | 6.022 (.082) | .053 (.007) |
| Sum-ANN | 6.009 (.084) | .057 (.007) |
| Bart-ANN | 6.021 (.083) | .053 (.007) |
| PV-QR | 6.021 (.085) | .053 (.008) |
| PV-LR | 6.097 (.082) | .029 (.007) |
| FS-LR | 6.101 (.081) | .028 (.007) |
| Sum-LR | 6.093 (.081) | .030 (.007) |
| LMS-Lin | 6.101 (.081) | .028 (.007) |
| LMS-Q | 6.091 (.083) | .031 (.007) |

*Note.* 30 repeated holdouts. LMS-Lin and LMS-Q converged in all 30 splits. LMS RMSE reflects the combined performance of latent-scale coefficient estimates applied to attenuated factor-score inputs (Section 3.4.1). Paired comparisons (Pearson $r \approx .93$–$.95$ from shared holdout splits) are reported both uncorrected and with the Nadeau and Bengio (2003) variance correction for overlapping training sets (correction factor 2.92 at 30 holdouts under the 80:20 split of Section 2.3, for which $n_2/n_1 = .25$). Sum-ANN exceeded PV-ANN, FS-ANN, and Bart-ANN (all $\Delta \geq +.0032$; all $p < .001$ uncorrected, all $p \leq .021$ corrected); PV-ANN was indistinguishable from FS-ANN ($\Delta = +.0009$; $p = .070$ uncorrected, $p = .524$ corrected) and from Bart-ANN ($\Delta = +.0007$; $p = .169$ uncorrected, $p = .632$ corrected). Full pairwise statistics are reported in Appendix H.1. The minimum detectable $|\Delta|$ at 80% power ($\alpha$ = .05, two-tailed) was .0013 uncorrected and .0039 corrected.

**Convergence diagnostics.** The Gibbs sampler converged across all 30 splits with quality comparable to the $\omega = .94$ simulation conditions, consistent with the moderate reliability in Study 2 (MH acceptance rate Mdn = .272, $\hat{R}_{\max}$ Mdn = 1.001, $\text{ESS}_{\min}$ Mdn = 3,247, FMI Mdn = .092, $B/W_K$ Mdn = 5.2).

**LMS-Q estimates.** LMS-Q detected a positive quadratic effect of Extraversion (standardized $\hat{\gamma}_{\text{quad1}} = .364$, $SD = .038$) and a negative Extraversion × Emotional Stability interaction (standardized $\hat{\gamma}_{\text{int12}} = -.208$, $SD = .038$), consistent with the positive curvature observed in Fig. 3. Full coefficient estimates are reported in Appendix H.2.

**Nonlinear increment.** The ANN-based nonlinear increment ($\Delta R^2 = .025 \pm .007$ for both PV-ANN − PV-LR and FS-ANN − FS-LR) was approximately eight times the LMS-based increment (LMS-Q − LMS-Lin: $\Delta R^2 = .003 \pm .003$). The gap is one of specification rather than of learner flexibility: PV-QR, a full second-order model in the three latent variables, matches the ANN increment (CV $R^2$ .053 against PV-LR's .029, versus PV-ANN's .054), whereas LMS-Q admits two of the six second-order terms such a model contains.

The CV $R^2$ pattern matched prediction (ii) of Remark 1. Paired $t$-tests across the 30 repeated holdouts, corrected for the variance understatement induced by overlapping training sets (Nadeau and Bengio 2003), left PV-ANN, FS-ANN and Bart-ANN statistically indistinguishable (Table 7 Note). What that establishes is the absence of a PV advantage, which is what the falsification check of Section 3.6 asks of it; it is consistent with prediction (ii) without being diagnostic of it, since all three inputs come from the same estimated measurement model and would coincide here whether that model is correct or misspecified. PV-ANN's nominal edge over FS-ANN (+.0009) points in the direction the theory does not permit but is nonsignificant before correction and clearly so after it.

Sum-ANN exceeded all three by +.0032, a margin that is statistically reliable after correction but amounts to about 6% of the CV $R^2$ itself and comes from a single dataset, so it is not interpreted as a finding here; the direction is nonetheless the one reported by Robitzsch (2026) for sum scores under unidimensional misspecification (see also McNeish 2023; Widaman and Revelle 2024), and it would locate any failure in a premise of Remark 1—correct measurement specification—rather than in the $\rho^k$ theory. Section 6.3.2 sets out the simulation that would test it. The cross-study difference in L2 grouping (Section 2.4) does not account for the pattern: rerunning Study 2 under Study 1's all-parameter regularization, with holdout splits, sampler seeds and hyperparameters held fixed, moved every ANN model's CV $R^2$ by at most .0002 and left the ordering intact.

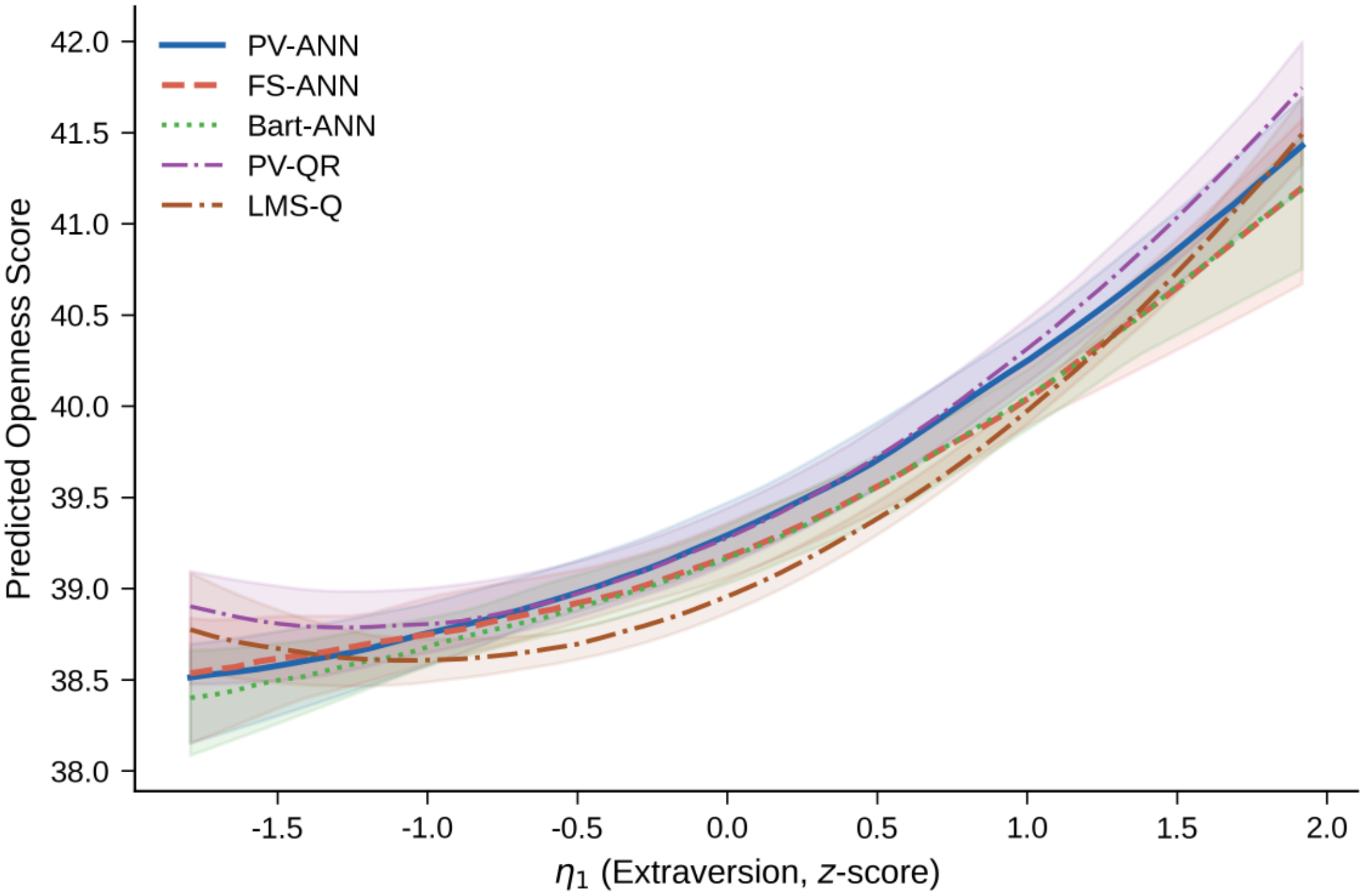

**Fig. 3** Extraversion → Openness partial dependence plot (Study 2: IPIP Big Five, $N = 10{,}000$, 30 repeated holdouts). Predicted Openness score as a function of $\eta_1$ (Extraversion), computed as Friedman's (2001) partial dependence function: for each of 50 grid values of $\eta_1$ (2.5th–97.5th percentile range), the prediction is averaged over 500 randomly sampled test observations, marginalizing over the empirical joint distribution of $(\eta_2, \eta_3)$. The $x$-axis is the $z$-scored regression factor score of Extraversion, used as a common coordinate; coordinate placement across model bases ($z$-matching) is detailed in Appendix H.3. Shaded regions represent pointwise 2.5/97.5 empirical percentile bands across the 30 repeated holdouts

All five models captured a predominantly increasing relationship with positive curvature over the displayed range, $\eta_1 \in [-1.8, +1.9]$. PV-ANN, FS-ANN, and Bart-ANN were predominantly non-decreasing across the displayed range, whereas LMS-Q and PV-QR exhibited a shallow local minimum in the negative range reflecting their parametric quadratic form (LMS-Q near its analytic vertex at $\eta_1 \approx -1$; Appendix H.3). The PV-ANN and FS-ANN curves were indistinguishable over the lower half of the range (median gap $+0.02$ points) and separated only slightly in the upper half ($+0.19$ points, against a residual standard deviation of 6.0), each median lying inside the other's percentile band throughout. The direction of that separation—PV-ANN above FS-ANN where the function is steepest—is the one the recovery argument predicts, and re-expressing the curves on the Section 3.6 coordinate raises it to $+0.26$ points, the change the two conventions imply (Appendix H.3). At either value it stays below a twentieth of the residual standard deviation, so it is not read as evidence either way. The convergence of PV-ANN and LMS-Q on the dominant increasing pattern supports the two-stage workflow (Section 6.2.1): PV-ANN reveals an unknown functional form, after which a parametric specification consistent with that form can be applied for confirmatory inference.

# 6 Discussion

## 6.1 Key Findings and Theoretical Implications

The four research questions map onto the theory as follows.

**RQ 1–2: recovery and its scaling.** PV-ANN exceeded point-estimate-based ANNs in nonlinear function shape recovery in all nonlinear conditions, with near-Oracle recovery ratios, closing roughly four fifths of the distance between FS-ANN and a learner that receives the true $\boldsymbol{\eta}$ (Section 4.3). The advantage scales with unreliability as the $\rho^k$ law requires: at $N = 1{,}000$ it is .064 (quadratic) and .038 (sigmoid) at $\omega = .80$ against .019 and .010 at $\omega = .94$, and equalizing ensemble size preserves that increase while all but removing the advantage where no nonlinear component is attenuated (Section 4.4). The score-type invariance of the recovery limit (Appendix A.2) was borne out empirically by Bart-ANN: although slightly superior to FS-ANN, both point-estimate-based scores were substantively inferior to PV-ANN across all nonlinear conditions. Because Bart-ANN uses population-parameter-based scores, its substantive inferiority to PV-ANN also indicates that the HP-selection design choice had a small effect relative to the score-type effect.

**RQ 3: nonlinear amplification.** The amplification difference was positive in all 18 conditions with CIs excluding zero, and at $\omega = .80$ the nonlinear DGPs (+.062 quadratic, +.031 sigmoid) exceeded the linear-DGP baseline (+.013) on both ensemble conventions—the $k \geq 2$ versus $k = 1$ contrast of Section 4.5, consistent with the attenuation law rather than an independent test of it. The LMS benchmark situates this result in the exploration–confirmation workflow: under correct specification, LMS-Q retains a parametric efficiency advantage (.040–.042), which shrinks by 80–86% under misspecification, and Fig. 2 shows that a symmetric quadratic form structurally cannot reproduce the asymmetric sigmoid residual structure. The complementary-relationship claim rests jointly on this quantitative reduction and the qualitative residual evidence.

**RQ 4: transfer and boundary conditions.** Study 2 served as a boundary probe and returned a predicted null and a predicted qualitative signal. The null: PV-ANN, FS-ANN, and Bart-ANN were statistically indistinguishable in CV $R^2$, the metric on which sufficiency forbids a PV advantage; Section 5.3 sets out why this parity is consistent with prediction (ii) without being diagnostic of it, and it is established only against effects above the corrected minimum detectable difference of .0039. The signal: the PDP recovered positive curvature converging with the LMS-Q parametric estimates, demonstrating the exploratory function of the framework on field data. Sum-ANN's edge of $+.003$ is about 6% of the metric and comes from a single dataset; it is noted but not read as a third result (Section 5.3).

**One scaling condition, one candidate boundary condition, and one premise.** (a) *Reliability (scaling condition):* the advantage is expected at all reliability levels but becomes substantive at approximately $\omega \lesssim .90$. The threshold is an interpolation anchored at the two studied levels—the population ceilings imply advantages of .001–.005 at $\omega = .94$ (Appendix C.3) against .018–.053 at $\omega = .80$ (Table 5)—with the crossover placed by the $\rho^k$ arithmetic; Study 1 establishes both endpoints. (b) *Signal strength (candidate boundary condition):* PV's absolute advantage is $(1 - \rho^k) \cdot \mathrm{Var}(f(\boldsymbol{\eta}))$; when $\mathrm{Var}(f(\boldsymbol{\eta}))$ is a small fraction of $\mathrm{Var}(Y)$, the recovery margin can be offset by PV's between-imputation variance overhead. The Study 2 case ($R^2_{\text{total}} \approx .054$) suggests an exploratory threshold of $R^2_{\text{total}} \gtrsim .10$; the latent-level threshold may be somewhat higher given DV-side measurement error (Section 5.1). (c) *Measurement specification (a premise, not an empirical finding):* both the recovery mechanism of Section 1.1 and the sufficiency argument of Remark 1 presume a correctly specified measurement model, so where it is misspecified the bias in $\widehat{\boldsymbol{\theta}}$ propagates to PVs drawn from $p(\boldsymbol{\eta} \mid \boldsymbol{X}, Y, \widehat{\boldsymbol{\theta}})$, whereas sum scores do not rely on measurement model estimation at all. This premise is checkable on the fitted model before any outcome is consulted, which is why Section 6.2.1 makes it a gate rather than a finding, and it needs no empirical corroboration here. Condition (b), by contrast, was suggested by the single case of Study 2 and is hypothesis-generating, requiring systematic verification (Section 6.3.2). When (b) or (c) is violated, sum-score-based approaches may be competitive with or superior to PV-ANN.

## 6.2 Relation to Existing Approaches and Practical Implications

PV-ANN combines two requirements that existing approaches have not simultaneously satisfied: non-specification of the nonlinear form and integration of measurement uncertainty via posterior sampling (Section 1.2). Its interpretation tool is the partial dependence plot (PDP; Friedman 2001), whereas LatentNN typically relies on internal network interpretation, the Bartlett-score approach on the fitted nonparametric curve, and LMS/SAM on the estimated structural coefficients. Because PDPs average over the marginal distribution of the remaining predictors, they can misrepresent effects when predictors are strongly correlated (Apley and Zhu 2020); the latent correlations here are modest (.30 in Study 1, .01–.37 in Study 2), but accumulated local effects would be the safer display at higher correlations.

### *6.2.1 Practical Application Guidelines*

**Recommended workflow.** PV-ANN is most effective as part of a two-stage analysis strategy: PDPs from PV-ANN identify the existence and shape of nonlinear patterns in the exploration stage, after which a parametric model consistent with the identified shape (LMS, SAM, or PV-QR) is used for confirmatory estimation.

Selecting the parametric form from the data and then testing it on those same data is a post-selection inference problem: the confirmatory $p$-values and confidence intervals do not retain their nominal level, because the specification was chosen with the outcome in view. The workflow therefore requires one of three provisions—splitting the sample so that the form is identified on one part and estimated on the other, confirming on an independent sample, or reporting the confirmatory stage as descriptive rather than as a test. Study 2 takes the third of these; a confirmatory claim was not made and none is available from a single application. This is a property of exploration followed by confirmation in general, not of PV-ANN; it is the cost of the workflow the framework recommends.

**Prediction stage.** Under a Gaussian measurement model the test-stage posterior of Section 2.3 is available in closed form, and practitioners should draw from it directly rather than reimplement the reduced Gibbs sampler, which is retained here only for extensibility to non-Gaussian measurement models (Section 6.3.3). The saving is the dominant cost of the prediction stage.

**Decision branches.** (a) For exploratory recovery of nonlinear function shape, PV-ANN is recommended when the conditions in Section 6.1 are met ($\omega \lesssim .90$, sufficient signal strength, adequately specified measurement model). At $\omega \geq .94$ the advantage at $N = 1{,}000$ is .010–.019, and .003–.012 once the point-estimate learners are given the same ensemble size (Table 4 panel (b)), which is limited relative to computational cost. Because the recovery advantage is bought by assuming the measurement model (Section 1.1), fit should be checked before the framework is applied and not only afterwards: where the measurement model falls below conventional cutoffs, the PV advantage is not guaranteed and a sum-score learner should be reported alongside PV-ANN as a reference, as Study 2 does. A practical minimum of $N \geq 1{,}000$ is recommended, with the post-burn-in length extended when reliability is low (Section 4.1). At $N \leq 300$, both ANN overfitting and Gibbs sampler instability limit applicability. (b) For minimizing RMSE, FS-ANN is competitive in both efficiency and accuracy. (c) For confirmatory testing, LMS/SAM is optimal when the form is theoretically predictable; otherwise, the exploration–confirmation workflow is recommended, with sample size planning supported by the model-implied simulation-based approach (Irmer et al. 2024b) in the powerNLSEM package (Irmer et al. 2024a).

**HP selection.** A grid search over hidden nodes $\{4,8,16\} \times \lambda_{L2}$ $\{10^{-4}, 10^{-3}, 10^{-2}\}$ is recommended (including 4 for applied contexts with potentially weaker signals), with selection based on PV-ANN's hold-out RMSE from 20 pilot replications.

**Convergence failure protocol.** Seed replacement (up to five attempts) is the quality-control gate used here (Section 3.5); where it does not suffice, extend the post-burn-in length (15,000 → 30,000), which on its own eliminated failures in the worst diagnostic condition tested, with results unchanged within MCSE (Section 4.1; Appendix C.6). Extension is the only step for which evidence is reported.

## 6.3 Limitations and Future Directions

### *6.3.1 Technical Limitations Within the Present Design*

**MCMC convergence.** The elevated failure rate under $\omega = .80$ did not affect the qualitative conclusions (Section 4.1) but remains a practical cost. Reparameterizing the conditioning model's design matrix with orthogonal polynomials is the cheapest available reparameterization, though it was not tested here; structural alternatives are noted in Appendix F.

**Approximate congeniality of the conditioning model.** Under the present conditions the second-order conditioning model did not substantively degrade $R^2_{\text{recovery}}$ (Section 4.2), but posterior distortion may be amplified under more complex nonlinear structures. A nonparametric conditioning model (e.g., BART; Chipman et al. 2010) is a possible mitigation.

**Prediction-stage draw count.** Step C retains one draw per posterior parameter set (Section 2.3), so the prediction integrates measurement uncertainty over $M = 5$ points. Averaging several draws within each $\boldsymbol{\theta}^{(m)}$ would approximate the same conditional expectation more precisely at negligible cost and move the RMSE comparison of Section 4.3 toward exact parity. Prediction (ii) of Remark 1 forbids a PV advantage on that metric but not the removal of the deficit, so this is a refinement of the estimator rather than a revision of the prediction.

**Ensemble size asymmetry.** Equalizing ensemble size removes the arithmetic part of the asymmetry, for the Oracle (Table C5) and for the point-estimate comparators (Table 4 panel (b), Table C9) alike, but a qualitative difference remains between PV-ANN's between-imputation diversity and the pure initialization diversity of the others.

### *6.3.2 Unmanipulated Factors*

Two follow-up simulations are needed to validate the boundary conditions in Section 6.1. The first varies overall signal strength ($R^2_{\text{pop}} \in \{.10, .30, .50\}$) to test the exploratory threshold of condition (b), since the present design's single high-signal condition substantially exceeds typical $R^2$ values in personality and educational psychology; the nonlinear share $\Delta R^2_{\text{NL}} / R^2_{\text{pop}}$ should be manipulated alongside it rather than left to covary with the choice of DGP, the present design spanning .00, .05, and .31 against applied cases at the lower end. The second manipulates the degree and type of measurement model misspecification jointly with signal strength, to locate the point at which PV quality degrades enough for sum scores to become competitive; because the recovery advantage is bought by assuming the measurement model (Section 1.1), this is the more consequential of the two. Sample size at $N \leq 300$ and $N \geq 5{,}000$, latent correlations $\geq .60$, non-normal latent distributions, and $P > 3$ latent variables warrant systematic investigation but are of secondary priority. The $\rho^k$ attenuation in Appendix A depends on normality, but by the variance inequality $\mathrm{Var}(E[f \mid \boldsymbol{X}]) \leq \mathrm{Var}(f)$, the qualitative direction holds regardless of distribution.

#### *6.3.3 Possible Extensions*

**Ordinal measurement model.** Replacing the measurement model block of the Gibbs sampler with an ordinal probit model enables PV extraction for ordinal indicators, contingent on an efficient sampler.

**Bartlett-score-based nonparametric regression.** Extending the $\rho^k$ analysis to Grønneberg and Irmer's (2024) Bartlett-score nonparametric regression—where the distortion mechanism differs qualitatively (noise inflation; Appendix A.2)—would systematically compare the finite-sample bias-variance trade-offs of the two approaches.

**Comparison with joint optimization approaches.** Direct simulation comparison against LatentNN (Ting 2026) requires a principled contrast across $\sigma^2_{\text{input}}$ specification strategies (oracle, estimated, misspecified) and is deferred to future research (Section 1.2).

### Data and Code Availability

Data, materials, and code are available in Harvard Dataverse (https://doi.org/10.7910/DVN/KD61HA). The Study 2 raw data are publicly available from the Open-Source Psychometrics Project (https://openpsychometrics.org/_rawdata/).

### Acknowledgments

During preparation of this manuscript, the authors used DeepL to translate the initial draft and improve English readability. The authors reviewed and edited the output and take full responsibility for the final content.

### Funding

No funding was received for this study.

### Competing Interests

The authors declare no competing interests.

# Appendices

## Appendix A. Derivation of $\rho^k$ Attenuation

### A.1 Regression Factor Scores

When $\eta \sim N(0,1)$ and the regression factor score is $\tilde{\eta} = \rho\eta + \sqrt{\rho(1-\rho)}\,\epsilon^*$, $\epsilon^* \sim N(0,1)$, $\epsilon^* \perp \eta$, then $\tilde{\eta} \sim N(0,\rho)$ and the joint distribution of $(\eta, \tilde{\eta})$ is bivariate normal. The conditional distribution satisfies $E[\eta \mid \tilde{\eta}] = \tilde{\eta}$ and $\eta \mid \tilde{\eta} \sim N(\tilde{\eta}, 1-\rho)$: the regression score is the projection of $\eta$ onto the observed-information $\sigma$-algebra, and variance shrinkage is manifested in $\mathrm{Var}(\tilde{\eta}) = \rho < 1$. Let $u \equiv \tilde{\eta}/\sqrt{\rho}$, $a = \sqrt{\rho}$, $b = \sqrt{1-\rho}$, so that $\eta = au + b\zeta$ with $\zeta \perp u$.

**Key property.** If $u, \zeta$ are independent standard normal and $\eta = au + b\zeta$ with $a^2 + b^2 = 1$, then

$$E[H_k(\eta) \mid u] = a^k H_k(u) \qquad \text{(A1)}$$

This is derived from the generating function of Hermite polynomials (Withers 2000): from $\sum_{k=0}^{\infty} H_k(\eta)\, t^k/k! = \exp(\eta t - t^2/2)$, substituting $\eta = au + b\zeta$ yields $\exp(\eta t - t^2/2) = \exp(au\,t - a^2t^2/2) \cdot \exp(b\zeta\,t - b^2t^2/2)$. Taking the conditional expectation given $u$ eliminates the $\zeta$-dependent factor, because for $\zeta \sim N(0,1)$ with $\zeta \perp u$, $E[\exp(b\zeta\,t - b^2t^2/2) \mid u] = 1$. The remaining factor $\exp(au\,t - a^2t^2/2) = \sum_{k=0}^{\infty} a^k\, H_k(u)\, t^k/k!$, and matching coefficients of $t^k$ yields Equation (A1). Equation (A1) is the classical eigenrelation of the bivariate-normal (Mehler) kernel (Lancaster 1957); the contribution here is its application to factor-score-based prediction and the score-type invariance established in Appendix A.2. Substituting $a = \sqrt{\rho}$,

$$E[H_k(\eta) \mid \tilde{\eta}] = \rho^{k/2}\, H_k\!\left(\frac{\tilde{\eta}}{\sqrt{\rho}}\right) \qquad \text{(A2)}$$

**Variance ratio.** Taking the variance of both sides of Equation (A2), $\mathrm{Var}(E[H_k(\eta) \mid \tilde{\eta}]) = \rho^k \cdot \mathrm{Var}(H_k(u)) = \rho^k \cdot k!$, and $\mathrm{Var}(H_k(\eta)) = k!$, so

$$\frac{\mathrm{Var}(E[H_k(\eta) \mid \tilde{\eta}])}{\mathrm{Var}(H_k(\eta))} = \rho^k \qquad \text{(A3)}$$

This proves that prediction based on regression factor scores captures only $\rho^k$ times the variance of the $k$th-order nonlinear component.

### A.2 Bartlett Factor Scores

The Bartlett score is $\tilde{\eta}_B = \eta + \delta$, where $\delta \perp \eta$ and $\mathrm{Var}(\delta) = 1/\rho - 1$ ($\delta$ is a linear function of the measurement error $\boldsymbol{\epsilon}$ in $\boldsymbol{X} = \boldsymbol{\Lambda}\eta + \boldsymbol{\epsilon}$). Thus $E[\tilde{\eta}_B \mid \eta] = \eta$ and $\mathrm{Var}(\tilde{\eta}_B) = 1/\rho > 1$. Deriving the conditional distribution: $\mathrm{Cov}(\eta, \tilde{\eta}_B) = 1$, $\mathrm{Var}(\tilde{\eta}_B) = 1/\rho$, so

$$E[\eta \mid \tilde{\eta}_B] = \rho\, \tilde{\eta}_B \qquad \text{(A4)}$$

and $\eta \mid \tilde{\eta}_B \sim N(\rho\,\tilde{\eta}_B,\, 1-\rho)$. Defining $u_B \equiv \sqrt{\rho}\,\tilde{\eta}_B$, then $u_B \sim N(0,1)$ and $\eta = \sqrt{\rho}\,u_B + \sqrt{1-\rho}\,\zeta^{(B)}$ ($\zeta^{(B)} \perp u_B$). Applying the same Hermite generating function technique,

$$E[H_k(\eta) \mid \tilde{\eta}_B] = \rho^{k/2}\, H_k(\sqrt{\rho}\,\tilde{\eta}_B) \qquad \text{(A5)}$$

Consequently, $\mathrm{Var}(E[H_k(\eta) \mid \tilde{\eta}_B])/\mathrm{Var}(H_k(\eta)) = \rho^k$, identical to Equation (A3). As discussed in Section 1.1, both score types are invertible linear transformations of the jointly sufficient statistic $\boldsymbol{\Lambda}'\boldsymbol{\Psi}^{-1}\boldsymbol{X}$, so $\sigma(\tilde{\eta}) = \sigma(\tilde{\eta}_B)$ and the information about $f(\eta)$ extractable under the optimal predictor is identical.

**General linear scores.** Neither derivation uses the particular form of the weights. Let $s = \boldsymbol{w}'\boldsymbol{X}$ be any linear composite with $\rho_s \equiv \text{Corr}^2(\eta, s) > 0$. Under the Gaussian measurement model $(\eta, s)$ is bivariate normal, so writing $u_s$ for the standardized $s$ gives $\eta = \sqrt{\rho_s}\, u_s + \sqrt{1-\rho_s}\,\zeta$ with $\zeta \perp u_s$, and Equation (A1) applies verbatim, yielding $\text{Var}(E[H_k(\eta) \mid s])/\text{Var}(H_k(\eta)) = \rho_s^k$. Equations (A3) and (A5) are the special case $\rho_s = \rho$. The sum score is bounded by its own, smaller coefficient of determination: in the single-factor case $\rho_{\text{sum}}$ equals $\omega$, which is $.800$ at the low-reliability level of the present design against $\rho = .818$ for the regression and Bartlett scores (the corresponding diagonals of the correlated-factor matrix are $.805$ and $.822$; Appendix C.3). The $\rho^k$ law is therefore a statement about how much of the $k$th-order signal a linear score retains, with the score entering only through its coefficient of determination. The slight empirical superiority of Bart-ANN (Section 4.3) is not a difference in the information the two scores carry about $\eta$, which Equation (A5) shows to be identical. In the single-factor case the training-set standardization of Section 3.4 makes the two inputs numerically identical ($\tilde{\eta} = \rho\, \tilde{\eta}_B$ exactly); with $P = 3$ correlated factors they remain nearly collinear (column-wise correlation $\approx .997$) but the regression score mixes a small cross-factor component through $\boldsymbol{\Phi}$ that the Bartlett score does not. The residual edge is of that size and is not interpreted further.

### A.3 Amplification Ratio Reference Benchmark

In the single-factor setting, for a pure $k$th-order component $f(\eta) = H_k(\eta)$, the capturable variance ratio of the factor-score-based optimal prediction is $\rho^k$ (Equation (A3)). Because the theoretical capture ratio under PV variance preservation is 1, the absolute advantage conferred by switching to PVs is $1 - \rho$ for the linear case and $1 - \rho^k$ for the $k$th-order case. The ratio of nonlinear to linear absolute advantage is

$$\frac{1-\rho^k}{1-\rho} = 1 + \rho + \rho^2 + \cdots + \rho^{k-1} \qquad \text{(A6)}$$

which equals $1 + \rho$ for $k = 2$. At $\rho \approx .818$ ($\omega = .80$), this is approximately 1.82; at $\rho \approx .949$ ($\omega = .94$), approximately 1.95. Extension to the correlated-factor case via the coefficient of determination matrix $\boldsymbol{R}^2 = \boldsymbol{\Phi\Lambda'\Sigma}^{-1}\boldsymbol{\Lambda\Phi}$ (Skrondal and Laake 2001) is summarized in Appendix C.3 below, where the accuracy of the univariate $\rho$ approximation in the present simulation is examined.

**Two coordinate conventions.** Equations (A3) and (A6) are statements about variance ratios. The quantity $R^2_{\text{recovery}}$ measures something else: the fit of the estimated function to the true function on the latent grid. Which power of $\rho_s$ appears depends on how a model's input axis is labelled, and both conventions occur in this study, for different purposes. This paragraph is the single statement of the relation; Sections 1.1, 3.6 and 5.3 and Appendix H.3 refer to it.

*Standardized coordinate.* A model fitted to a $z$-scored linear score returns $h(u) = \sum_k c_k\, \rho_s^{k/2} H_k(u)$ at standardized coordinate $u$, so read against $u$ the $k$th-order coefficient is attenuated by $\rho_s^{k/2}$. This is the convention of any display whose axis is the standardized score, including the partial dependence plot of Section 5.3.

*Latent coordinate.* Under the evaluation rule of Section 3.6 the same learner is evaluated at $u = \sqrt{\rho_s}\,\eta_g$; since the leading term of $H_k(\sqrt{\rho_s}\,\eta)$ is $\rho_s^{k/2} H_k(\eta)$, the $k$th Hermite coefficient of the recovered function is $c_k \rho_s^k$. This is the convention of $R^2_{\text{recovery}}$ and of the $\rho^k$ bound of Section 1.1, and the same exponent therefore governs both the capturable variance ratio of Equation (A3) and the attenuation of the recovered coefficient.

The two differ only in the labelling of each model's input axis, so a curve archived under one can be re-expressed on the other without refitting; Appendix H.3 reports that re-expression for Fig. 3. A case whose true latent value is $\eta$ obtains $u = \sqrt{\rho_s}\,\eta$ in expectation, so the re-expression stretches the latent axis of a score-based model by $1/\sqrt{\rho_s}$ and leaves a PV-based model ($\rho_s = 1$) unchanged.

$R^2_{\text{recovery}}$ penalizes that attenuation quadratically: for a pure $k$th-order component the loss is $(1-\rho^k)^2$ times the component's share of the function variance on the grid, and the nonlinear-to-linear ratio of losses is $(1+\rho)^2 \approx 3.31$ at $\rho = .818$ rather than the $1+\rho \approx 1.82$ of Equation (A6). Equation (A6) is accordingly read in this study as a prediction about the *ordering* of PV advantages across nonlinear orders (RQ 3), not as a calibrated numerical target for Diff or Amp.

Because a PV-based model ($\rho_s = 1$) is unattenuated under either convention while every score-based model is attenuated under both, the choice rescales the gaps but preserves the ordering of models on $R^2_{\text{recovery}}$; Section 3.6 adopts the latent coordinate because that is the coordinate on which the estimand is defined.

## Appendix B. IPIP Big Five Analytic Sample Characteristics

The analytic sample for Study 2 ($N = 10{,}000$) was a random subsample drawn from $N_{\text{eligible}} = 603{,}322$ respondents who met the criteria of complete responses, valid range, and IPC = 1 (IPC is the number of records from the same IP address in the dataset; the value 1 retains a single record per IP address, screening duplicate and shared-network submissions, per the dataset codebook), out of the full publicly available dataset ($N_{\text{total}} = 1{,}015{,}341$).

**Table B1** Descriptive statistics for the IPIP Big Five analytic sample ($N = 10{,}000$)

| Scale | *M* | *SD* | $\alpha$ |
|---|---|---|---|
| Extraversion | 29.31 | 9.20 | .901 |
| Emot. Stability | 29.30 | 8.60 | .873 |
| Agreeableness | 37.59 | 7.43 | .846 |
| Openness (DV) | 39.28 | 6.21 | .802 |

*Note.* $\alpha$ = Cronbach's $\alpha$. Computed after recoding reverse-scored items. Table B1 reports $\alpha$ following descriptive-statistics conventions; CFA-based composite reliability $\omega$ (McDonald 1999) is reported in Table 6. In the present data, the difference between $\alpha$ and $\omega$ was at most .003 (Agreeableness .846 vs. .849; Emotional Stability .873 vs. .877; Openness .802 vs. .802), reflecting the limited heterogeneity of factor loadings characteristic of the congeneric model. The $\omega$ for Openness, used for the attenuation correction in Section 5.1, comes from a one-factor congeneric CFA on its 10 items; that model fits poorly (CFI = .658), but $\alpha$ and $\omega$ agree to three decimals, so the reliability estimate does not depend on the unidimensional specification. Because the analytic sample was randomly drawn from the eligible sample ($N_{\text{eligible}} = 603{,}322$; seed = 20260316; Section 5.1), differences in descriptive statistics fall within sampling error. The raw data are publicly accessible from the Open-Source Psychometrics Project (https://openpsychometrics.org/_rawdata/).

## Appendix C. Computational Time, Population Verification, and Sensitivity Analyses

### C.1 Computational Time by Condition

The median wall time per replication was 32–55 seconds under the $\omega = .94$ condition and 43–119 seconds under the $\omega = .80$ condition. The discrepancy between mean and median in the $\omega = .80$ condition is attributable to the distribution of re-run replications triggered by the convergence quality-control gate. For the sigmoid $\omega = .80$, $N = 500$ condition, the median (119.3 s) exceeded the mean (94.0 s) because the majority of replications (75.2%) underwent one or more retries, recording high wall times, whereas a minority converged on the first attempt, pulling the mean downward with shorter wall times. Under the $\omega = .94$ condition, no re-runs occurred, and the two statistics coincided. Detailed values by condition are available in the deposited package.

### C.2 Selected Hyperparameters, FMI, and Variance Decomposition (All 18 Conditions)

**Table C1** Selected hyperparameters, FMI, and variance decomposition (all 18 conditions)

| **Condition** | $N$ | **Hidden nodes** | $\lambda_{L2}$ | **FMI Mdn** | $B/W_K$ | **ΔRMSE ($M = 3$ vs. 5)** | $\Delta R^2$ **($M = 3$ vs. 5)** |
|---|---|---|---|---|---|---|---|
| Linear, $\omega = .94$ | 500 | 16 | 1e-02 | .155 | 2.4 | .0019 | .0024 |
| Linear, $\omega = .94$ | 1,000 | 8 | 1e-04 | .157 | 1.8 | .0019 | .0019 |
| Linear, $\omega = .94$ | 2,000 | 16 | 1e-04 | .154 | 4.0 | .0016 | .0013 |
| Linear, $\omega = .80$ | 500 | 16 | 1e-03 | .238 | 6.9 | .0057 | .0051 |
| Linear, $\omega = .80$ | 1,000 | 16 | 1e-04 | .236 | 9.4 | .0051 | .0028 |
| Linear, $\omega = .80$ | 2,000 | 8 | 1e-04 | .234 | 6.8 | .0054 | .0020 |
| Quadratic, $\omega = .94$ | 500 | 16 | 1e-04 | .157 | 2.7 | .0027 | .0044 |
| Quadratic, $\omega = .94$ | 1,000 | 16 | 1e-04 | .153 | 3.7 | .0021 | .0019 |
| Quadratic, $\omega = .94$ | 2,000 | 16 | 1e-04 | .148 | 4.4 | .0020 | .0014 |
| Quadratic, $\omega = .80$ | 500 | 16 | 1e-03 | .223 | 8.2 | .0063 | .0049 |
| Quadratic, $\omega = .80$ | 1,000 | 16 | 1e-03 | .221 | 11.8 | .0065 | .0030 |
| Quadratic, $\omega = .80$ | 2,000 | 16 | 1e-04 | .220 | 14.0 | .0057 | .0024 |
| Sigmoid, $\omega = .94$ | 500 | 16 | 1e-02 | .157 | 2.4 | .0016 | .0019 |
| Sigmoid, $\omega = .94$ | 1,000 | 16 | 1e-03 | .155 | 3.1 | .0016 | .0013 |
| Sigmoid, $\omega = .94$ | 2,000 | 16 | 1e-03 | .153 | 4.1 | .0014 | .0012 |
| Sigmoid, $\omega = .80$ | 500 | 8 | 1e-02 | .234 | 4.6 | .0055 | .0039 |
| Sigmoid, $\omega = .80$ | 1,000 | 8 | 1e-02 | .231 | 6.1 | .0044 | .0029 |
| Sigmoid, $\omega = .80$ | 2,000 | 16 | 1e-02 | .232 | 18.0 | .0045 | .0012 |

*Note.* This table is referenced in Section 4.1. See Section 2.4 for the definition of $B/W_K$. The full range is 1.8–4.4 for $\omega = .94$ and 4.6–18.0 for $\omega = .80$. The $B/W_K = 18.0$ for the sigmoid $\omega = .80$, $N = 2{,}000$ condition is the largest across all conditions, reflecting that when low reliability

and large sample size combine, measurement uncertainty (between-PV variability) dominates algorithmic uncertainty (between-initialization variability).

### C.3 Population Parameter Verification

**Multivariate $\rho$ verification.** Diagonal elements of the coefficient of determination matrix $\boldsymbol{R}^2$: (.949, .949, .949) under the $\omega = .94$ condition and (.822, .822, .822) under the $\omega = .80$ condition. The corresponding univariate $\rho$ (Equation (2)): .949 and .818, respectively. The differences were .000 and .004.

**Correlated-factor extension of $\rho^k$ attenuation** (referenced from Appendix A.3 above). In the single-factor case, the capturable variance ratio of the $k$th-order Hermite component under factor-score-based prediction is $\rho^k$ (Equation (A3)). With several correlated factors ($P > 1$, $\boldsymbol{\Phi} \neq \boldsymbol{I}$), the shrinkage structure is described by the coefficient of determination matrix

$$\boldsymbol{R}^2 = \boldsymbol{\Phi}\boldsymbol{\Lambda}'\boldsymbol{\Sigma}^{-1}\boldsymbol{\Lambda}\boldsymbol{\Phi} \qquad \text{(C1)}$$

where $\boldsymbol{\Sigma} = \boldsymbol{\Lambda}\boldsymbol{\Phi}\boldsymbol{\Lambda}' + \boldsymbol{\Psi}$ is the model-implied observed covariance matrix (Skrondal and Laake 2001, Eq. 13). The diagonal elements of $\boldsymbol{R}^2$ generalize the univariate $\rho$ to each latent dimension under the joint multivariate measurement model, and off-diagonal elements describe cross-factor shrinkage. Because $k$th-order multivariate Hermite components also depend on this cross-factor structure, the univariate $\rho^k$ benchmark is an approximation.

Under the homogeneous latent correlation of .30 used here, the off-diagonal cross-factor shrinkage was small and the univariate $\rho^k$ approximation deviates from the diagonal by at most .004 (values above). Stronger latent correlations or heterogeneous loading patterns would enlarge the deviation, motivating systematic investigation of $\boldsymbol{\Phi}$ as a follow-up factor (Section 6.3.2).

**Population-optimal score-based recovery.** Evaluating the optimal predictor $E[f(\boldsymbol{\eta}) \mid \boldsymbol{s}]$ on the grid under the rule of Section 3.6 gives the ceiling attainable by any point-estimate-based learner, and $1 -$ ceiling is the PV advantage that $\rho^k$ attenuation alone implies, the Oracle's ceiling being exactly $1.000$. Table 5 reports the $\omega = .80$ comparison; at $\omega = .94$ the ceilings are $.999$ (linear), $.995$ (quadratic) and $.997$ (sigmoid), implying advantages of $.001$, $.005$ and $.003$. Under the Gaussian measurement model the conditional expectation is a fixed Gaussian quadrature applied at shifted means, so these are computed rather than simulated; `scripts/export_population_ceiling.py` in the archived code emits them and checks the quadrature against the closed form available for the linear and quadratic DGPs, which it reproduces to machine precision. The ceilings for regression and Bartlett scores are numerically identical in every condition, which is the counterpart of the score-type invariance proved in Appendix A.2; the sum score, whose coefficient of determination is smaller, sits below both ($.978$, $.938$, $.961$ at $\omega = .80$).

**DGP $R^2$ decomposition.** Population $R^2_{\text{pop}} = .500$. $\Delta R^2_{\text{NL}}$: DGP 1 .000, DGP 2 .155, DGP 3 .025.

**Conditioning model $R^2$.** DGP 1 1.000, DGP 2 1.000, DGP 3 .968.

### C.4 Convergence Sensitivity Analysis

**Table C2** Differences in PV-ANN results by inclusion/exclusion of non-converged replications ($\omega$ = .80 conditions)

| Condition | $N$ | Total $R_{\text{rep}}$ | Converged $R_{\text{rep}}$ | Fail | ΔRMSE | $\Delta R^2$ |
|---|---|---|---|---|---|---|
| Linear, $\omega$ = .80 | 500 | 250 | 198 | 52 | −.0005 | −.0024 |
| Linear, $\omega$ = .80 | 1,000 | 500 | 466 | 34 | −.0001 | −.0003 |
| Linear, $\omega$ = .80 | 2,000 | 250 | 246 | 4 | −.0001 | −.0001 |
| Quadratic, $\omega$ = .80 | 500 | 250 | 166 | 84 | +.0002 | −.0033 |
| Quadratic, $\omega$ = .80 | 1,000 | 500 | 444 | 56 | −.0006 | −.0002 |
| Quadratic, $\omega$ = .80 | 2,000 | 250 | 238 | 12 | −.0004 | −.0006 |
| Sigmoid, $\omega$ = .80 | 500 | 250 | 139 | 111 | +.0025 | +.0008 |
| Sigmoid, $\omega$ = .80 | 1,000 | 500 | 390 | 110 | −.0003 | +.0014 |
| Sigmoid, $\omega$ = .80 | 2,000 | 250 | 220 | 30 | −.0000 | +.0001 |

*Note.* Δ = results from all replications − results from converged replications only. All conditions under $\omega = .94$ had Fail = 0 and are therefore omitted. The maximum $|\Delta|$ was .003 for both RMSE and $R^2$, well within the MCSE range.

### C.5 Pilot Replication Exclusion Sensitivity Analysis

**Table C3** Differences in PV-ANN results when excluding the 20 pilot replications

| DGP | $\omega$ | $N$ | ΔRMSE | $\Delta R^2$ |
|---|---|---|---|---|
| Linear | .94 | 500 | +.0009 | −.0005 |
| Linear | .94 | 1,000 | +.0004 | +.0001 |
| Linear | .94 | 2,000 | −.0010 | +.0000 |
| Linear | .80 | 500 | +.0012 | −.0016 |
| Linear | .80 | 1,000 | +.0002 | +.0001 |
| Linear | .80 | 2,000 | −.0002 | +.0001 |
| Quadratic | .94 | 500 | −.0008 | −.0005 |
| Quadratic | .94 | 1,000 | −.0000 | −.0000 |
| Quadratic | .94 | 2,000 | −.0002 | −.0001 |
| Quadratic | .80 | 500 | −.0000 | −.0003 |
| Quadratic | .80 | 1,000 | +.0003 | +.0004 |
| Quadratic | .80 | 2,000 | −.0003 | −.0003 |
| Sigmoid | .94 | 500 | +.0002 | −.0004 |
| Sigmoid | .94 | 1,000 | +.0001 | −.0002 |
| Sigmoid | .94 | 2,000 | −.0000 | +.0002 |

| DGP | $\omega$ | $N$ | ΔRMSE | $\Delta R^2$ |
|---|---|---|---|---|
| Sigmoid | .80 | 500 | +.0008 | −.0014 |
| Sigmoid | .80 | 1,000 | −.0001 | +.0001 |
| Sigmoid | .80 | 2,000 | −.0001 | +.0003 |

*Note.* Δ = results after excluding the pilot replications − full results, recomputed from the archived run-of-record. The 20 pilot replications of each condition (rep 0–19) are those whose hold-out RMSE selected that condition's hyperparameters (Section 3.5), so they overlap the main run. The maximum $|\Delta|$ across all 18 conditions was **.0012** for RMSE and **.0016** for $R^2_{\text{recovery}}$, both within the MCSE range, indicating that optimistic bias due to pilot–main-run overlap in hyperparameter selection was not practically meaningful. Values are produced by the `pilot_exclusion` table routine and checked by `python scripts/verify_manuscript_claims.py` in the deposited package. **Correction.** The version of this table in the manuscript under review reported nine rows and different values, because its generating routine selected replications by convergence-retry status rather than by pilot index; that rule also excluded the $\omega = .94$ conditions, in which no replication is ever re-run. The routine has been corrected, the table above is the analysis the text describes, and the conclusion is unchanged and more strongly supported.

### C.6 Chain Length Sensitivity Analysis

To examine whether convergence failures observed under the $\omega = .80$ condition (Table C6, Appendix C.8) could be resolved by extending chain length, the post-burn-in length was compared at 15K (default) and 30K (extended) in the worst-convergence condition (sigmoid, $\omega = .80$, $N = 1{,}000$; 100 replications).

**Table C4** Chain length sensitivity analysis: sigmoid, $\omega = .80$, $N = 1{,}000$

| Condition | $\hat{R}_{\max}$ Mdn | $\text{ESS}_{\min}$ Mdn | Fail (%) | RMSE | $R^2_{\text{recovery}}$ |
|---|---|---|---|---|---|
| 15K (default) | 1.004 | 222 | 23.0 | .600 ($\pm$.003) | .936 ($\pm$.002) |
| 30K (extended) | 1.002 | 360 | 0.0 | .601 ($\pm$.003) | .937 ($\pm$.002) |

*Note.* Fail = proportion of replications with $\hat{R} \geq 1.05$ or ESS $< 200$. Values in parentheses are $SD$.

The 30K extension eliminated failures, reducing the Fail rate from 23.0% to 0.0%, while increasing the $\text{ESS}_{\min}$ median from 222 to 360 (a 62% increase). The differences in RMSE and $R^2_{\text{recovery}}$ were .001 and .001, respectively, both within the MCSE range, suggesting that convergence failures under the 15K default length had only a marginal impact on results. Meanwhile, the pattern in which RMSE slightly increased (.600 → .601) while $R^2_{\text{recovery}}$ slightly improved (.936 → .937) under the 30K extension is consistent with the $R^2_{\text{recovery}}$–RMSE trade-off discussed in Section 4.3. These results are the evidence behind the chain-extension step of the convergence failure protocol in Section 6.2.1.

### C.7 Oracle $K = 15$ Ensemble Equalization Analysis

Oracle-ANN was expanded to $K = 15$ across all 18 conditions, matching PV-ANN's $M \times K = 15$ (6,000 replications, 1.41 hours; condition-specific hyperparameters from Table C1). Appendix C.11 applies the same equalization to the point-estimate comparators.

**Table C5** Oracle-ANN $K = 3$ vs. $K = 15$ comparison and ratio to PV-ANN ($R^2_{\text{recovery}}$)

| DGP | $\omega$ | $N$ | Oracle $K = 3$ | Oracle $K = 15$ | Δ | PV-ANN | PV / $\text{Ora}_{K15}$ |
|---|---|---|---|---|---|---|---|
| Linear | .94 | 500 | .9593 | .9696 | +.010 | .9666 | .997 |
| Linear | .94 | 1,000 | .9715 | .9822 | +.011 | .9812 | .999 |
| Linear | .94 | 2,000 | .9858 | .9902 | +.004 | .9895 | .999 |
| Linear | .80 | 500 | .9547 | .9677 | +.013 | .9439 | .975 |
| Linear | .80 | 1,000 | .9757 | .9832 | +.007 | .9715 | .988 |
| Linear | .80 | 2,000 | .9820 | .9897 | +.008 | .9860 | .996 |
| Quadratic | .94 | 500 | .8923 | .9016 | +.009 | .8948 | .992 |
| Quadratic | .94 | 1,000 | .9357 | .9424 | +.007 | .9411 | .999 |
| Quadratic | .94 | 2,000 | .9547 | .9595 | +.005 | .9603 | 1.001† |
| Quadratic | .80 | 500 | .8805 | .8941 | +.014 | .8691 | .972 |
| Quadratic | .80 | 1,000 | .9362 | .9421 | +.006 | .9320 | .989 |
| Quadratic | .80 | 2,000 | .9576 | .9610 | +.003 | .9550 | .994 |
| Sigmoid | .94 | 500 | .9207 | .9319 | +.011 | .9283 | .996 |
| Sigmoid | .94 | 1,000 | .9412 | .9507 | +.010 | .9478 | .997 |
| Sigmoid | .94 | 2,000 | .9590 | .9642 | +.005 | .9612 | .997 |
| Sigmoid | .80 | 500 | .9188 | .9321 | +.013 | .9136 | .980 |
| Sigmoid | .80 | 1,000 | .9359 | .9450 | +.009 | .9373 | .992 |
| Sigmoid | .80 | 2,000 | .9573 | .9610 | +.004 | .9524 | .991 |

*Note.* Δ = Oracle $K = 15$ − Oracle $K = 3$. † = sampling variability within MCSE (.0009). Across all 18 conditions, Oracle $K = 15$ exceeded $K = 3$, with Δ ranging from +.003 to .014. PV / $\text{Ora}_{K15}$ did not exceed 1 beyond MCSE in any condition (one condition at 1.001, within the MCSE of .0009), indicating that ensemble-size equalization is required for an unbiased Oracle benchmark. All Oracle-benchmark comparisons in Sections 4.3–4.5 are therefore reported on the $K = 15$ basis.

### C.8 Gibbs Sampler Convergence Diagnostic Summary

**Table C6** Gibbs sampler convergence diagnostic summary (all 18 conditions)

| Condition | $N$ | MH Mdn | $\hat{R}_{\text{max}}$ Mdn | $\text{ESS}_{\text{min}}$ Mdn | Rerun rate (%) | Fail (%) |
|---|---|---|---|---|---|---|
| Linear, $\omega$ = .94 | 500 | .270 | 1.001 | 1,119 | 0.0 | 0.0 |
| Linear, $\omega$ = .94 | 1,000 | .271 | 1.001 | 1,246 | 0.0 | 0.0 |

| Condition | $N$ | MH Mdn | $\widehat{R}_{\mathbf{max}}$ Mdn | $\text{ESS}_{\text{min}}$ Mdn | Rerun rate (%) | Fail (%) |
|---|---|---|---|---|---|---|
| Linear, $\omega$ = .94 | 2,000 | .271 | 1.001 | 1,277 | 0.0 | 0.0 |
| Linear, $\omega$ = .80 | 500 | .276 | 1.006 | 228 | 55.6 | 20.8 |
| Linear, $\omega$ = .80 | 1,000 | .277 | 1.005 | 242 | 40.8 | 6.8 |
| Linear, $\omega$ = .80 | 2,000 | .278 | 1.005 | 254 | 30.0 | 1.6 |
| Quadratic, $\omega$ = .94 | 500 | .270 | 1.001 | 1,040 | 0.0 | 0.0 |
| Quadratic, $\omega$ = .94 | 1,000 | .271 | 1.001 | 1,097 | 0.0 | 0.0 |
| Quadratic, $\omega$ = .94 | 2,000 | .271 | 1.001 | 1,183 | 0.0 | 0.0 |
| Quadratic, $\omega$ = .80 | 500 | .273 | 1.006 | 216 | 68.4 | 33.6 |
| Quadratic, $\omega$ = .80 | 1,000 | .275 | 1.005 | 233 | 47.2 | 11.2 |
| Quadratic, $\omega$ = .80 | 2,000 | .275 | 1.005 | 245 | 39.2 | 4.8 |
| Sigmoid, $\omega$ = .94 | 500 | .271 | 1.001 | 884 | 0.0 | 0.0 |
| Sigmoid, $\omega$ = .94 | 1,000 | .271 | 1.001 | 950 | 0.0 | 0.0 |
| Sigmoid, $\omega$ = .94 | 2,000 | .271 | 1.001 | 997 | 0.0 | 0.0 |
| Sigmoid, $\omega$ = .80 | 500 | .275 | 1.005 | 203 | 75.2 | 44.4 |
| Sigmoid, $\omega$ = .80 | 1,000 | .277 | 1.005 | 221 | 62.2 | 22.0 |
| Sigmoid, $\omega$ = .80 | 2,000 | .278 | 1.004 | 223 | 56.4 | 12.0 |

*Note.* MH Mdn = Metropolis-Hastings acceptance rate median. Rerun rate = proportion rerun at the quality control gate. Fail (%) = percentage of replications unconverged after a maximum of 5 retries (Fail / $R_{\text{rep}} \times 100$). See Section 4.1 for discussion.

### C.9 PV Bias by True $\eta_1$ Interval

**Table C7** Mean PV bias (PV mean − true value) by true $\eta_1$ interval ($N = 1{,}000$, marker-variable scale)

| DGP | $\omega$ | $[-2.0, -1.5)$ | $[-1.5, -1.0)$ | $[-1.0, -0.5)$ | $[-0.5, 0.0)$ | $[0.0, +0.5)$ | $[+0.5, +1.0)$ | $[+1.0, +1.5)$ | $[+1.5, +2.0]$ | max \|bias\| |
|---|---|---|---|---|---|---|---|---|---|---|
| Linear | .94 | +.093 | +.068 | +.037 | +.014 | −.013 | −.041 | −.066 | −.093 | .093 |
| Linear | .80 | —[a] | +.240 | +.142 | +.044 | −.046 | −.146 | −.245 | −.400 | .400 |
| Quadratic | .94 | +.098 | +.064 | +.042 | +.012 | −.012 | −.039 | −.066 | −.102 | .102 |
| Quadratic | .80 | +.280 | +.236 | +.142 | +.050 | −.046 | −.139 | −.246 | −.305 | .305 |
| Sigmoid | .94 | +.100 | +.070 | +.041 | +.014 | −.012 | −.040 | −.066 | −.097 | .100 |
| Sigmoid | .80 | +.197 | +.248 | +.145 | +.051 | −.049 | −.149 | −.248 | −.440 | .440 |

*Note.* Bias = PV mean − true $\eta_1$, averaged across replications ($R_{\text{rep}} = 500$). Values are on the marker-variable scale ($\eta_1 = \lambda_1^* \cdot \eta_1^*$; $\lambda_1^* = .85$ for $\omega = .94$ and $\lambda_1^* = .60$ for $\omega = .80$). The monotone pattern (positive bias for $\eta_1 < 0$, negative for $\eta_1 > 0$) reflects shrinkage toward the prior mean. Bias magnitude at $\omega = .80$ is approximately 3–4 times that at $\omega = .94$, consistent with the greater posterior uncertainty under lower measurement reliability. [a]The extreme bin $[-2.0, -1.5)$ in the linear $\omega = .80$ condition contained fewer than 5 observations in many replications, producing unstable estimates; this value is suppressed. See Section 4.2.

**C.10 Estimation-Setting Sensitivity for the Residual ANN Gap**

Section 4.6 reports that every ANN estimator in Fig. 2, Oracle-ANN included, leaves a residual gap against the true function, and attributes it to the estimation setting shared by all models rather than to measurement error. Table C8 reports the runs behind that attribution. Oracle-ANN is fitted under DGP 3, $N = 1{,}000$, $K = 15$, on the data of the first 100 main-run replications, so that measurement error is absent and the gap is isolated from it. Hidden width, the L2 penalty, and the early-stopping budget are varied one at a time from the values selected for this condition (8 nodes, $\lambda_{L2} = .01$, patience 10, 500 steps), and then jointly.

**Table C8** Oracle-ANN estimation-setting sensitivity, DGP 3, $\omega = .80$, $N = 1{,}000$

| **Variation** | **Hidden** | $\boldsymbol{\lambda_{L2}}$ | **Patience / steps** | $\boldsymbol{R^2_{\text{recovery}}}$ | **Δ paired (*SE*)** | **Tail[a]** | **Center[a]** |
|---|---|---|---|---|---|---|---|
| Default | 8 | .01 | 10 / 500 | .9427 | — | .0936 | .0958 |
| Wider network | 32 | .01 | 10 / 500 | .9504 | +.0077 (.0011) | .0784 | .0835 |
| Wider network | 128 | .01 | 10 / 500 | .9320 | −.0107 (.0022) | .1020 | .0963 |
| Weaker L2 | 8 | .0001 | 10 / 500 | .9435 | +.0008 (.0004) | .0926 | .0922 |
| Relaxed early stop | 8 | .01 | 50 / 5,000 | .9490 | +.0063 (.0006) | .0868 | .0850 |
| Relaxed early stop | 8 | .01 | 100 / 20,000 | .9501 | +.0075 (.0007) | .0860 | .0833 |
| All three jointly | 32 | .0001 | 100 / 20,000 | .9531 | +.0104 (.0014) | .0622 | .0635 |

*Note.* $R_{\text{rep}} = 100$ per row, sharing data with the corresponding main-run replications; the default row reproduces the archived Oracle-ANN $K = 15$ value for this condition (Table C5) to within its Monte Carlo error. Δ paired is the within-replication difference in $R^2_{\text{recovery}}$ against the default setting and is the inferential quantity here, the marginal MCSE ($.0018$–$.0033$) being larger than the paired standard errors by an order of magnitude. [a]Mean absolute residual of the median prediction over the 162 grid points with $|\eta_1| \geq 1.5$ (Tail) and over the remaining 567 (Center). Values are produced by `python scripts/run_budget_capacity.py` in the deposited package.

Two readings follow. The L2 penalty is not the operative constraint: weakening it by two orders of magnitude at the default width and budget moves $R^2_{\text{recovery}}$ by $.0008$. Width and budget both matter but not separably, since widening to 32 nodes helps while widening to 128 hurts—at 500 gradient steps the larger network is not trained to its own optimum—so neither can be credited alone. No arm removes the gap, the best reducing it by about a third and moving the tail and central residuals in nearly the same proportion, which is the basis for the attribution in Section 4.6.

**C.11 Ensemble-Size Equalization for the Point-Estimate Comparators**

Appendix C.7 equalizes Oracle-ANN to $K = 15$ on the ground that ensemble-size equalization is required for an unbiased benchmark. The same asymmetry holds against the comparators that carry RQ 1–3: PV-ANN averages $M \times K = 15$ networks while FS-ANN, Sum-ANN and Bart-ANN average $K = 3$ (Table 3). Table C9 reports what those three gain when given the same fifteen.

The design is nested. Fifteen networks are trained per learner and the $K = 3$ arm is the mean of the first three of that same fifteen, so the two arms share their data, hyperparameters and initializations and the paired difference isolates ensemble size with no seed-stream confound. That confound is not hypothetical: two independent $K = 3$ streams evaluated on the same replications differed from each other by $+.0085$ ($SE$ $.0058$), which is the same order as the effect being measured. $R_{\text{rep}} = 100$ per condition, sharing data with the main run; the $K = 3$ arm reproduces the archived 500-replication values to within $.0064$ in all 54 cells (mean deviation $.0001$). Because the paired $\Delta$ is estimated an order of magnitude more precisely than either arm's mean, the equalized gaps below are formed as the archived value minus $\Delta$ rather than from the 100-replication means.

**Table C9** Ensemble-size equalization of FS-ANN and its effect on the PV advantage

| DGP | $\omega$ | $N$ | $\Delta$ ($K = 15 - K = 3$) | $SE$ | PV − FS as reported | PV − FS equalized |
|---|---|---|---|---|---|---|
| Linear | .94 | 500 | +.0106 | .0014 | .0115 | .0009 |
| Linear | .94 | 1,000 | +.0096 | .0010 | .0135 | .0039 |
| Linear | .94 | 2,000 | +.0049 | .0007 | .0069 | .0020 |
| Linear | .80 | 500 | +.0078 | .0014 | .0116 | .0038 |
| Linear | .80 | 1,000 | +.0077 | .0015 | .0201 | .0124 |
| Linear | .80 | 2,000 | +.0060 | .0009 | .0278 | .0218 |
| Quadratic | .94 | 500 | +.0131 | .0045 | .0281 | .0150 |
| Quadratic | .94 | 1,000 | +.0064 | .0019 | .0187 | .0123 |
| Quadratic | .94 | 2,000 | +.0036 | .0009 | .0156 | .0120 |
| Quadratic | .80 | 500 | +.0077 | .0036 | .0622 | .0545 |
| Quadratic | .80 | 1,000 | +.0044 | .0020 | .0643 | .0599 |
| Quadratic | .80 | 2,000 | +.0038 | .0015 | .0594 | .0556 |
| Sigmoid | .94 | 500 | +.0104 | .0021 | .0155 | .0051 |
| Sigmoid | .94 | 1,000 | +.0072 | .0014 | .0102 | .0030 |
| Sigmoid | .94 | 2,000 | +.0047 | .0010 | .0070 | .0023 |
| Sigmoid | .80 | 500 | +.0118 | .0023 | .0377 | .0259 |
| Sigmoid | .80 | 1,000 | +.0079 | .0020 | .0379 | .0300 |
| Sigmoid | .80 | 2,000 | +.0035 | .0010 | .0296 | .0261 |

*Note.* $R_{\text{rep}} = 100$ per condition. $\Delta$ is the within-replication difference in $R^2_{\text{recovery}}$ between the $K = 15$ and nested $K = 3$ ensembles. Sum-ANN and Bart-ANN were equalized in the same run

and gave $\Delta$ of $.0025$–$.0144$ and $.0021$–$.0121$ respectively, so the gain is not specific to the regression score. Values are produced by `python scripts/run_fs_ensemble_equalization.py` in the deposited package.

Three readings follow; the $N = 1{,}000$ equalized values themselves are in Table 4 panel (b) and are not restated here.

First, $\Delta$ is positive in every cell for every learner, so the asymmetry is real and the panel (a) gaps overstate the contribution of plausible values by that amount.

Second, the overstatement is strongly unequal across conditions, and unequal in the direction the theory requires. In the nonlinear $\omega = .80$ conditions, which carry RQ 1, the gap loses 6–12% (quadratic) and 12–31% (sigmoid) of its size; elsewhere it loses 22–92%, most of that under the linear DGP and at $\omega = .94$, where $\rho^k$ attenuation of a nonlinear component is absent (linear DGP) or small ($\omega = .94$). What survives equalization is concentrated where the attenuation law predicts it should be, which is why Section 4.4 reads the reliability scaling as essentially unchanged ($.046 \rightarrow .048$ quadratic, $.028 \rightarrow .027$ sigmoid) on a high-reliability baseline that has dropped nearly to zero. For RQ 3, subtracting each condition's Sum-ANN $\Delta$ leaves Diff positive in all six $N = 1{,}000$ cells, at $+.004$ to $+.054$ against $+.012$ to $+.062$ as reported.

Third, the ceiling identity of Section 4.3 tightens. At $\omega = .80$, $N = 2{,}000$ the equalized gaps are $.022$, $.056$ and $.026$ against ceiling-implied advantages of $.018$, $.053$ and $.033$ (Table 5); the largest departure falls from $.010$ to $.007$, and the linear-DGP excess from $.0098$ to $.0038$, so most of what Section 4.5 would otherwise attribute to a DGP-independent mechanism was ensemble size.

## Appendix D. Results Tables and Figure

### D.1 Convergence of the LMS Benchmark

**Table D1** LMS convergence diagnostics summary (see Section 4.1)

| Model | Total conditions | Convergence rate (%) | Warnings/negative variances |
|---|---|---|---|
| LMS-Lin | 18 | 100 | 0 |
| LMS-Q | 18 | 100 | 0 |

*Note.* Conditions = 3 DGPs × 2 $\omega$ × 3 $N$. All replications ($R_{\text{rep}}$) within each condition converged. Mplus estimation settings: `TYPE = RANDOM; ALGORITHM = INTEGRATION; INTEGRATION = MONTECARLO(5000); CONVERGENCE = 0.00005; MITERATIONS = 1000; MCONVERGENCE = 0.001`. Study 2 used the same settings except `INTEGRATION = MONTECARLO(2000)` (Appendix H.2). Replications with convergence failure are excluded from results.

### D.2 Full Results by Sample Size

Tables 4–5 report $N = 1{,}000$ ($R_{\text{rep}} = 500$). The following tables report all 18 conditions × 11 models, adding $N = 500$ and $N = 2{,}000$ ($R_{\text{rep}} = 250$); Table 4 is the corresponding block of Table D2.

**Table D2** $R^2_{\text{recovery}}$: all 18 conditions

| DGP | $\omega$ | $N$ | Oracle | PV-ANN | FS-ANN | Sum-ANN | Bart-ANN | PV-QR | LMS-Lin | LMS-Q | PV-LR | FS-LR | Sum-LR |
|---|---|---|---|---|---|---|---|---|---|---|---|---|---|
| Linear | .94 | 500 | .9696 | .9666 | .9551 | .9550 | .9564 | .9187 | .9782 | .9649 | .9791 | .9827 | .9823 |
| Linear | .94 | 1,000 | .9822 | .9812 | .9677 | .9678 | .9686 | .9626 | .9896 | .9824 | .9903 | .9907 | .9902 |
| Linear | .94 | 2,000 | .9902 | .9895 | .9826 | .9814 | .9825 | .9829 | .9950 | .9919 | .9953 | .9949 | .9945 |
| Linear | .80 | 500 | .9677 | .9439 | .9323 | .9234 | .9298 | .8168 | .9682 | .9466 | .9577 | .9683 | .9646 |
| Linear | .80 | 1,000 | .9832 | .9715 | .9514 | .9472 | .9514 | .9245 | .9850 | .9750 | .9821 | .9750 | .9710 |
| Linear | .80 | 2,000 | .9897 | .9860 | .9582 | .9547 | .9582 | .9676 | .9924 | .9874 | .9919 | .9783 | .9745 |
| Quadratic | .94 | 500 | .9016 | .8948 | .8667 | .8658 | .8733 | .9172 | .6369 | .9641 | .6382 | .6398 | .6397 |
| Quadratic | .94 | 1,000 | .9424 | .9411 | .9224 | .9203 | .9245 | .9632 | .6490 | .9826 | .6498 | .6504 | .6502 |
| Quadratic | .94 | 2,000 | .9595 | .9603 | .9447 | .9427 | .9448 | .9818 | .6551 | .9913 | .6557 | .6554 | .6551 |
| Quadratic | .80 | 500 | .8941 | .8691 | .8069 | .8059 | .8142 | .7889 | .6232 | .9338 | .6179 | .6270 | .6248 |
| Quadratic | .80 | 1,000 | .9421 | .9320 | .8677 | .8616 | .8726 | .9232 | .6454 | .9718 | .6443 | .6386 | .6358 |
| Quadratic | .80 | 2,000 | .9610 | .9550 | .8956 | .8898 | .8977 | .9659 | .6534 | .9849 | .6531 | .6439 | .6413 |
| Sigmoid | .94 | 500 | .9319 | .9283 | .9128 | .9145 | .9165 | .8884 | .9188 | .9370 | .9191 | .9210 | .9206 |
| Sigmoid | .94 | 1,000 | .9507 | .9478 | .9376 | .9370 | .9382 | .9317 | .9268 | .9539 | .9272 | .9287 | .9284 |
| Sigmoid | .94 | 2,000 | .9642 | .9612 | .9542 | .9523 | .9548 | .9530 | .9317 | .9626 | .9320 | .9321 | .9317 |
| Sigmoid | .80 | 500 | .9321 | .9136 | .8759 | .8730 | .8848 | .7805 | .9062 | .9161 | .8898 | .9051 | .9007 |
| Sigmoid | .80 | 1,000 | .9450 | .9373 | .8994 | .8960 | .9023 | .8962 | .9215 | .9453 | .9171 | .9107 | .9064 |
| Sigmoid | .80 | 2,000 | .9610 | .9524 | .9228 | .9177 | .9229 | .9375 | .9304 | .9597 | .9280 | .9138 | .9093 |

*Note.* Oracle = Oracle-ANN ($K = 15$). LMS-Lin = linear structural model only; LMS-Q = specification including quadratic interaction. See Section 3.4.1 for discussion of the structural asymmetry in $R^2_{\text{recovery}}$ computation between LMS and PV-ANN and the scale mismatch in LMS RMSE.

**Table D3** $R^2_{\text{recovery},w}$: all 18 conditions

| DGP | $\omega$ | $N$ | Oracle | PV-ANN | FS-ANN | Sum-ANN | Bart-ANN | PV-QR | LMS-Lin | LMS-Q | PV-LR | FS-LR | Sum-LR |
|---|---|---|---|---|---|---|---|---|---|---|---|---|---|
| Linear | .94 | 500 | .9691 | .9655 | .9555 | .9558 | .9569 | .9414 | .9737 | .9652 | .9765 | .9801 | .9797 |
| Linear | .94 | 1,000 | .9825 | .9811 | .9698 | .9702 | .9707 | .9729 | .9874 | .9829 | .9891 | .9895 | .9890 |
| Linear | .94 | 2,000 | .9901 | .9894 | .9838 | .9830 | .9838 | .9874 | .9940 | .9920 | .9948 | .9943 | .9940 |
| Linear | .80 | 500 | .9673 | .9414 | .9350 | .9256 | .9318 | .8729 | .9641 | .9509 | .9552 | .9659 | .9622 |
| Linear | .80 | 1,000 | .9830 | .9703 | .9544 | .9504 | .9545 | .9473 | .9828 | .9764 | .9809 | .9738 | .9698 |
| Linear | .80 | 2,000 | .9899 | .9852 | .9640 | .9617 | .9644 | .9773 | .9911 | .9879 | .9912 | .9776 | .9738 |
| Quadratic | .94 | 500 | .9204 | .9128 | .8941 | .8931 | .9005 | .9345 | .6820 | .9622 | .6875 | .6894 | .6892 |
| Quadratic | .94 | 1,000 | .9547 | .9523 | .9426 | .9423 | .9445 | .9698 | .7057 | .9815 | .7060 | .7066 | .7064 |

| DGP | $\omega$ | $N$ | Oracle | PV-ANN | FS-ANN | Sum-ANN | Bart-ANN | PV-QR | LMS-Lin | LMS-Q | PV-LR | FS-LR | Sum-LR |
|---|---|---|---|---|---|---|---|---|---|---|---|---|---|
| Quadratic | .94 | 2,000 | .9701 | .9694 | .9625 | .9616 | .9624 | .9852 | .7114 | .9904 | .7123 | .7120 | .7117 |
| Quadratic | .80 | 500 | .9157 | .8760 | .8506 | .8505 | .8564 | .8397 | .6738 | .9361 | .6681 | .6784 | .6759 |
| Quadratic | .80 | 1,000 | .9553 | .9381 | .9068 | .9016 | .9102 | .9400 | .6967 | .9718 | .6972 | .6906 | .6872 |
| Quadratic | .80 | 2,000 | .9711 | .9621 | .9328 | .9287 | .9345 | .9732 | .7099 | .9855 | .7099 | .6996 | .6964 |
| Sigmoid | .94 | 500 | .9260 | .9229 | .9077 | .9094 | .9114 | .9072 | .9099 | .9297 | .9127 | .9118 | .9111 |
| Sigmoid | .94 | 1,000 | .9454 | .9425 | .9331 | .9324 | .9332 | .9342 | .9200 | .9438 | .9217 | .9208 | .9203 |
| Sigmoid | .94 | 2,000 | .9587 | .9549 | .9491 | .9475 | .9493 | .9471 | .9253 | .9513 | .9262 | .9245 | .9238 |
| Sigmoid | .80 | 500 | .9279 | .9101 | .8733 | .8700 | .8792 | .8439 | .9004 | .9143 | .8919 | .8934 | .8886 |
| Sigmoid | .80 | 1,000 | .9408 | .9323 | .8971 | .8927 | .8998 | .9138 | .9153 | .9382 | .9145 | .8991 | .8942 |
| Sigmoid | .80 | 2,000 | .9549 | .9456 | .9182 | .9122 | .9186 | .9383 | .9245 | .9495 | .9236 | .9027 | .8977 |

*Note.* Oracle = Oracle-ANN ($K = 15$).

**Table D4** Hold-out RMSE: all 18 conditions

| DGP | $\omega$ | $N$ | Oracle | PV-ANN | FS-ANN | Sum-ANN | Bart-ANN | PV-QR | LMS-Lin | LMS-Q | PV-LR | FS-LR | Sum-LR |
|---|---|---|---|---|---|---|---|---|---|---|---|---|---|
| Linear | .94 | 500 | .6720 | .6846 | .6846 | .6873 | .6851 | .6873 | .6804 | .6825 | .6812 | .6794 | .6808 |
| Linear | .94 | 1,000 | .6715 | .6850 | .6855 | .6865 | .6849 | .6860 | .6810 | .6819 | .6829 | .6805 | .6819 |
| Linear | .94 | 2,000 | .6692 | .6826 | .6813 | .6826 | .6811 | .6828 | .6789 | .6795 | .6812 | .6788 | .6803 |
| Linear | .80 | 500 | .6729 | .7207 | .7155 | .7200 | .7155 | .7257 | .7107 | .7128 | .7176 | .7087 | .7126 |
| Linear | .80 | 1,000 | .6685 | .7176 | .7116 | .7168 | .7114 | .7196 | .7086 | .7096 | .7161 | .7078 | .7130 |
| Linear | .80 | 2,000 | .6699 | .7160 | .7104 | .7145 | .7102 | .7165 | .7077 | .7082 | .7150 | .7072 | .7114 |
| Quadratic | .94 | 500 | .6595 | .6757 | .6785 | .6808 | .6776 | .6636 | .7384 | .6587 | .7389 | .7376 | .7386 |
| Quadratic | .94 | 1,000 | .6502 | .6658 | .6662 | .6683 | .6657 | .6584 | .7343 | .6545 | .7353 | .7340 | .7349 |
| Quadratic | .94 | 2,000 | .6491 | .6656 | .6648 | .6662 | .6647 | .6591 | .7344 | .6553 | .7356 | .7342 | .7351 |
| Quadratic | .80 | 500 | .6600 | .7128 | .7108 | .7146 | .7106 | .7108 | .7540 | .6977 | .7575 | .7519 | .7544 |
| Quadratic | .80 | 1,000 | .6522 | .7101 | .7054 | .7106 | .7050 | .7077 | .7580 | .6954 | .7624 | .7575 | .7602 |
| Quadratic | .80 | 2,000 | .6454 | .7022 | .6963 | .7013 | .6963 | .6993 | .7533 | .6889 | .7580 | .7531 | .7559 |
| Sigmoid | .94 | 500 | .5608 | .5728 | .5726 | .5741 | .5726 | .5763 | .5726 | .5705 | .5742 | .5720 | .5732 |
| Sigmoid | .94 | 1,000 | .5584 | .5715 | .5708 | .5722 | .5709 | .5736 | .5738 | .5701 | .5754 | .5734 | .5747 |
| Sigmoid | .94 | 2,000 | .5583 | .5712 | .5698 | .5710 | .5696 | .5728 | .5742 | .5702 | .5757 | .5739 | .5751 |
| Sigmoid | .80 | 500 | .5630 | .6059 | .6017 | .6048 | .6010 | .6110 | .6014 | .5999 | .6074 | .5997 | .6025 |
| Sigmoid | .80 | 1,000 | .5601 | .6017 | .5967 | .6004 | .5968 | .6040 | .5976 | .5952 | .6038 | .5968 | .6001 |
| Sigmoid | .80 | 2,000 | .5570 | .6005 | .5941 | .5976 | .5942 | .6019 | .5966 | .5939 | .6032 | .5963 | .5995 |

*Note.* Oracle = Oracle-ANN ($K = 15$).

**Table D5** PV advantage decomposition (amplification): all 18 conditions

**(a) Sum Baseline**

| DGP | $\omega$ | $N$ | Lin | NL | Diff [95% CI] | Amp |
|---|---|---|---|---|---|---|
| Linear | .94 | 500 | −.0032 | .0115 | +.015 [.013, .017] | n.e. |
| Linear | .94 | 1,000 | .0001 | .0134 | +.013 [.012, .015] | n.e. |
| Linear | .94 | 2,000 | .0008 | .0081 | +.007 [.006, .009] | n.e. |
| Linear | .80 | 500 | −.0070 | .0205 | +.028 [.021, .034] | n.e. |
| Linear | .80 | 1,000 | .0111 | .0243 | +.013 [.011, .015] | 2.19 [1.94, 2.51] |
| Linear | .80 | 2,000 | .0175 | .0314 | +.014 [.012, .016] | 1.80 [1.68, 1.93] |
| Quadratic | .94 | 500 | −.0015 | .0290 | +.031 [.024, .039] | n.e. |
| Quadratic | .94 | 1,000 | −.0004 | .0208 | +.021 [.018, .024] | n.e. |
| Quadratic | .94 | 2,000 | .0005 | .0176 | +.017 [.015, .019] | n.e. |
| Quadratic | .80 | 500 | −.0069 | .0631 | +.070 [.061, .079] | n.e. |
| Quadratic | .80 | 1,000 | .0086 | .0703 | +.062 [.057, .066] | 8.19 [6.90, 10.08] |
| Quadratic | .80 | 2,000 | .0118 | .0652 | +.053 [.049, .058] | 5.55 [4.92, 6.35] |
| Sigmoid | .94 | 500 | −.0015 | .0138 | +.015 [.012, .019] | n.e. |
| Sigmoid | .94 | 1,000 | −.0012 | .0108 | +.012 [.010, .014] | n.e. |
| Sigmoid | .94 | 2,000 | .0002 | .0089 | +.009 [.007, .010] | n.e. |
| Sigmoid | .80 | 500 | −.0109 | .0405 | +.051 [.045, .058] | n.e. |
| Sigmoid | .80 | 1,000 | .0108 | .0413 | +.031 [.028, .033] | 3.84 [3.31, 4.63] |
| Sigmoid | .80 | 2,000 | .0187 | .0347 | +.016 [.014, .018] | 1.86 [1.74, 2.00] |

**(b) FS Baseline**

| DGP | $\omega$ | $N$ | Lin | NL | Diff [95% CI] | Amp |
|---|---|---|---|---|---|---|
| Linear | .94 | 500 | −.0036 | .0114 | +.015 [.013, .017] | n.e. |
| Linear | .94 | 1,000 | −.0004 | .0135 | +.014 [.013, .015] | n.e. |
| Linear | .94 | 2,000 | .0004 | .0069 | +.007 [.006, .008] | n.e. |
| Linear | .80 | 500 | −.0106 | .0117 | +.022 [.018, .027] | n.e. |
| Linear | .80 | 1,000 | .0071 | .0201 | +.013 [.011, .015] | 2.83 [2.38, 3.52] |
| Linear | .80 | 2,000 | .0136 | .0278 | +.014 [.013, .016] | 2.04 [1.88, 2.22] |
| Quadratic | .94 | 500 | −.0017 | .0281 | +.030 [.023, .037] | n.e. |
| Quadratic | .94 | 1,000 | −.0006 | .0187 | +.019 [.017, .022] | n.e. |
| Quadratic | .94 | 2,000 | .0003 | .0156 | +.015 [.014, .017] | n.e. |
| Quadratic | .80 | 500 | −.0091 | .0621 | +.071 [.062, .081] | n.e. |
| Quadratic | .80 | 1,000 | .0057 | .0643 | +.059 [.054, .063] | 11.27 [8.90, 15.53] |
| Quadratic | .80 | 2,000 | .0091 | .0594 | +.050 [.046, .056] | 6.49 [5.61, 7.68] |
| Sigmoid | .94 | 500 | −.0018 | .0155 | +.017 [.014, .021] | n.e. |

| DGP | ω | N | Lin | NL | Diff [95% CI] | Amp |
|---|---|---|---|---|---|---|
| Sigmoid | .94 | 1,000 | −.0015 | .0102 | +.012 [.010, .014] | n.e. |
| Sigmoid | .94 | 2,000 | −.0001 | .0070 | +.007 [.006, .008] | n.e. |
| Sigmoid | .80 | 500 | −.0153 | .0377 | +.053 [.047, .060] | n.e. |
| Sigmoid | .80 | 1,000 | .0064 | .0379 | +.031 [.029, .034] | 5.88 [4.64, 8.21] |
| Sigmoid | .80 | 2,000 | .0142 | .0296 | +.016 [.014, .017] | 2.09 [1.92, 2.29] |

*Note.* Diff = NL − Lin (primary metric, in $R^2_{\text{recovery}}$ units). 95% bootstrap CI (10,000 resamples). Amp = NL / Lin (secondary metric). n.e. = not estimable (|Lin| < .005 or Lin ≤ 0, rendering the ratio unstable). Negative Lin values appear in every $N = 500$ cell and in some $\omega = .94$ cells at larger $N$, where the between-imputation variance of PVs offsets the small bias reduction available to linear prediction. Consequently, the amplification ratio is estimable only in the $\omega = .80$ conditions with $N \geq 1{,}000$. Across all 18 conditions, Diff > 0 and the CI excluded zero, and in all six estimable cells the Amp CI excluded 1, so the preregistered ratio form supports the amplification hypothesis wherever it is estimable. Panel (b) restates the decomposition on the regression-factor-score baseline; the two baselines support the same conclusions (Section 3.6).

### D.3 Supplementary Contour Plot: Additional Models for Fig. 1

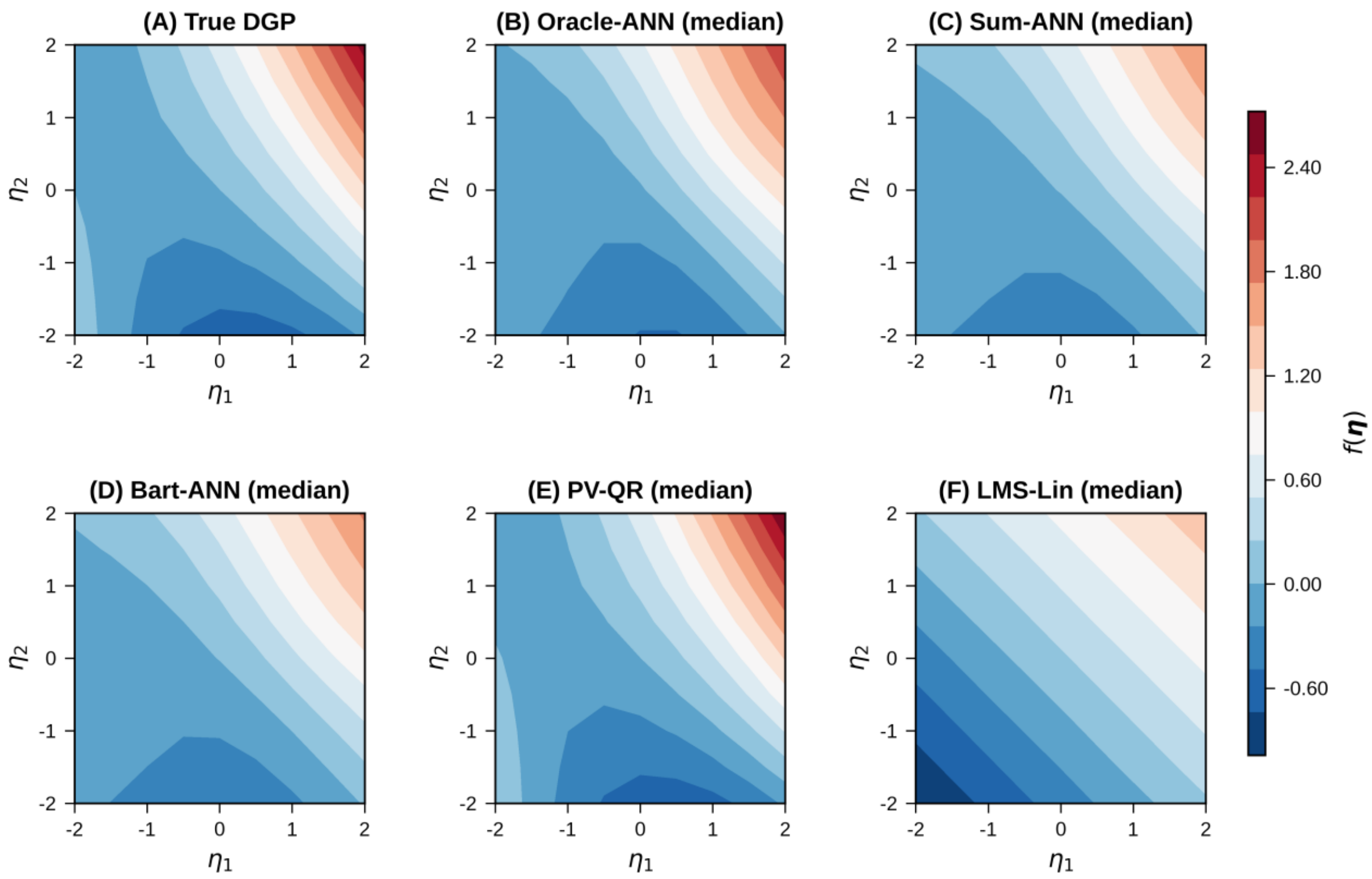

**Fig. D1** Supplementary contour plot for the DGP 2, $\omega = .80$, $N = 1{,}000$ condition: additional reference models

*Note.* Extends Fig. 1 with models not displayed therein. Contours show $f(\boldsymbol{\eta})$ over $(\eta_1, \eta_2)$ with $\eta_3 = 0$ held fixed; median of 500 replications. Panels: (A) True DGP (reproduced from Fig. 1 for reference), (B) Oracle-ANN, (C) Sum-ANN, (D) Bart-ANN, (E) PV-QR, (F) LMS-Lin. Model definitions follow Table 3 and Section 3.4. Oracle-ANN ($K = 15$) is the benchmark attainable under error-free input at the shared training budget of Section 2.4. Sum-ANN and Bart-ANN are alternative point-estimate-based ANNs; their visible variance-shrinkage attenuation parallels FS-ANN in Fig. 1. PV-QR combines PVs with a parametric quadratic predictor, and LMS-Lin is the linear-only LMS specification shown as a baseline. Color-scale range differs slightly from Fig. 1 because range is set independently for each figure to fit the displayed models' value ranges.

**Appendix E. Structural Asymmetry in the LMS–PV-ANN $\boldsymbol{R^2_{\text{recovery}}}$ Comparison**

Under DGP 2 the quadratic form of LMS-Q is exactly correct, and a correctly specified parametric model is more efficient than a nonparametric one: LMS-Q converges to $R^2_{\text{recovery}} \to 1$ as $N \to \infty$ while PV-ANN retains an irreducible approximation floor, and LMS-Q accordingly reached 102–107% of Oracle-ANN ($K = 15$) there, exceeding even a learner that sees the true $\boldsymbol{\eta}$. The computation works in the same direction: $R^2_{\text{recovery}}$ is obtained for LMS by substituting the estimated coefficients $\hat{\gamma}^*$ into the parametric form on the true $\eta^*$ grid, so that it reflects parameter estimation precision only, whereas for PV-ANN it applies the ensemble average $\frac{1}{MK} \sum_{m,k}$ $\hat{f}^{(m,k)}$ $(\boldsymbol{\eta}^*_g)$ to the same grid and so additionally reflects function approximation error.

The informative comparison is therefore DGP 3, where misspecification bias gives LMS-Q its own irreducible floor: the LMS-Q/Oracle ratio falls to 98.3–100.5% (99.8–100.3% for $N \geq 1{,}000$). What the quadratic form cannot reproduce there is the origin-asymmetric saturation of the $\eta_2 \cdot \text{sigmoid}(5\eta_1)$ term, which linearizes for $\eta_1 \gg 0$ and vanishes for $\eta_1 \ll 0$ (Section 3.2), against the origin-symmetric $\eta_1\eta_2$ interaction of LMS-Q.

LMS RMSE combines two scales: coefficients estimated on the latent scale are applied at test time to regression factor scores, which are attenuated, so the quadratic contribution enters with variance $\rho^2$ times the true value and the underprediction grows as $\rho$ falls. The common observed-score input basis serves the comparison, but LMS is a group-level estimation tool rather than an individual-prediction one and its RMSE should be read accordingly.

**Appendix F. Technical Details of MCMC Sampler Improvement Paths**

None of the options below was run, and none is part of the archived run-of-record; they are recorded as directions rather than as tested remedies. Orthogonalizing the conditioning model's design matrix improves the condition number of the $\boldsymbol{\gamma}$ posterior at the lowest cost, because it leaves the block structure unchanged. Three structural alternatives each carry a known limitation: a collapsed Gibbs sampler marginalizing $\boldsymbol{\gamma}$ is tractable under a normal prior but does not extend to nonlinear conditioning models; $(\boldsymbol{\gamma}, \boldsymbol{\eta})$ joint block Metropolis-Hastings suits $P = 3$ but degrades in acceptance rate as $P$ grows; and the No-U-Turn Sampler (Hoffman and Gelman 2014) scales in high dimensions but complicates implementation alongside the measurement model blocks $(\boldsymbol{\Lambda}, \boldsymbol{\Psi})$. The only step for which evidence is reported is extending the post-burn-in length (Section 4.1).

## Appendix G. Bias–Variance Decomposition of the PV–FS RMSE Gap

The directional divergence between $R^2_{\text{recovery}}$ and RMSE reported in Section 4.3 is explained by the following bias–variance decomposition. The approximation neglects Bias × variance cross-terms, which are small when the bias is approximately constant across observations within each replication:

$$\text{RMSE}^2_{\text{PV-ANN}} \approx \text{Bias}^2_{\text{PV}} + V_{\text{within}} + V_{\text{between}} \qquad \text{(G1)}$$

$$\text{RMSE}^2_{\text{FS-ANN}} \approx \text{Bias}^2_{\text{FS}} + V_{\text{within,FS}} + 0 \qquad \text{(G2)}$$

The direction of change for each term of PV-ANN relative to FS-ANN is as follows:

- $\text{Bias}^2$: substantially decreased with PVs ($\rho^k$ attenuation removed); includes $\rho^k$ attenuation with FS.
- $V_{\text{within}}$: comparable in both models (within-initialization variance under the $K$-ensemble).
- $V_{\text{between}}$: additional in PVs (between-imputation variance from posterior distribution sampling); zero in FS (single input).

Variance restoration corrects the systematic distortion of the function structure without necessarily improving point-estimate precision, which is why the principal contribution of PV-ANN lies in function shape recovery rather than in individual prediction accuracy. The $B/W_K$ ratio of Table C1 quantifies $V_{\text{between}}$ relative to $V_{\text{within}}$ across conditions.

## Appendix H. Supplementary Analyses for Study 2

### H.1 Full Pairwise Comparison Statistics

Table H1 reports the paired $t$-tests across the 30 repeated holdouts on the uncorrected scale; the Nadeau and Bengio (2003) variance-corrected values for the three focal ANN pairs, together with the corrected minimum detectable difference, are given in the Table 7 note (correction factor 2.92). The high between-model correlation from shared splits (Pearson $r \approx .93$–$.95$; paired $SD$s .0024–.0029 vs. marginal $SD$s .0073–.0075) accounts for the substantially smaller paired-test $SD$ relative to the marginal $SD$ of CV $R^2$ in Table 7.

**Table H1** Within-ANN-group paired comparisons of CV $R^2$ (30 repeated holdouts)

| Pair | $M(\Delta$ CV $R^2)$ | $p$ (two-tailed) |
|---|---|---|
| Sum-ANN − PV-ANN | +.0032 | < .001 |
| Sum-ANN − FS-ANN | +.0041 | < .001 |
| Sum-ANN − Bart-ANN | +.0039 | < .001 |
| PV-ANN − FS-ANN | +.0009 | .070 |
| PV-ANN − Bart-ANN | +.0007 | .169 |
| FS-ANN − Bart-ANN | ≤ .001 [a] | n.r. |

*Note.* Δ = mean of the within-holdout-pair differences across 30 splits, equivalent to $M_A - M_B$ from Table 7. All $p$ values in this table are uncorrected. n.r. = not reported. Paired-test $SD$s ranged .0024–.0029 across pairs. The minimum detectable $|\Delta|$ at 80% power ($\alpha = .05$, two-tailed) was approximately .0013 (uncorrected). [a]$|\Delta| \leq .001$, below the minimum detectable difference of .0013; the paired test is omitted because the design lacks power at this magnitude. Complete pairwise test results for all model pairs (including ANN-vs-LR and within-LR comparisons) are reported in `table-pairwise_tests.csv` in the deposited package.

Across ANN vs LR comparisons (e.g., PV-ANN − PV-LR), the observed Δ values fell in the .022–.029 range, far exceeding the detection floor, and all corresponding paired tests yielded $p < .001$.

**H.2 LMS-Q Coefficient Estimates**

The LMS-Q model `Y ON eta1 eta2 eta3 int12 quad1` was estimated in Mplus using `TYPE = RANDOM` (Section 3.4.1) across the 30 repeated holdouts; estimation settings were identical to Study 1 (Table D1 Note) except `INTEGRATION = MONTECARLO(2000)`. Coefficient estimates are summarized in Table H2.

**Table H2** LMS-Q structural coefficients (marker-variable scale; 30 repeated holdouts)

| Coefficient | Unstandardized $M$ ($SD$) | Standardized $M$ ($SD$) |
|---|---|---|
| $\eta_1$ | 0.836 (0.046) | .722 (.040) |
| $\eta_2$ | 0.458 (0.043) | .426 (.040) |
| $\eta_3$ | 0.600 (0.079) | .406 (.053) |
| $\eta_1\eta_2$ | −0.259 (0.047) | −.208 (.038) |
| $\eta_1^2$ | 0.488 (0.051) | .364 (.038) |

*Note.* $M$ and $SD$ across 30 splits. Standardized coefficients were computed using the standard deviations of the latent variables. The nonlinear increment associated with the quadratic and interaction terms was $\Delta R^2 = .003$ (LMS-Q − LMS-Lin; see Section 5.3). The two key coefficients reported in Section 5.3—the positive quadratic effect of Extraversion ($\hat{\gamma}_{\text{quad1}} = .364$, $SD = .038$) and the negative Extraversion × Emotional Stability interaction ($\hat{\gamma}_{\text{int12}} = -.208$, $SD = .038$)—are reproduced in this table for context. The LMS-Q vertex on the standardized scale, computed from these coefficients, is reported in Appendix H.3.

**H.3 Partial Dependence Plot: Additional Analyses**

This section provides detailed analyses underlying the partial dependence plot (PDP) reported in Fig. 3.

**Coordinate placement across model bases ($z$-matching).** The $x$-axis of Fig. 3 is the $z$-scored regression factor score of Extraversion, used as the common coordinate for all models. Inputs are placed by $z$-matching: each model receives individual conditional expectation (ICE) samples at the identical standardized coordinates in its own training $z$-space—FS $z$-space for FS-ANN, Bartlett $z$-space for Bart-ANN (on the same common grid), and each PV set's training $z$-space for PV-ANN, PV-QR, and PV-LR, averaged across the $M = 5$ sets. Because factor-score and PV coordinates are unit-scale standardizations of estimates of the same latent variable, $z$-matching places every model's input at the same standardized position, aligning models by standardized position rather than by latent value: the first of the two conventions of Appendix A.3. The released code enforces $z$-matching by default and retains the alternative latent-value placement (`transform = "raw_rescale"`) for sensitivity use. In the archived run Bart-ANN was stored on its own $\eta_1$ grid rather than the common one; realigning it moves its curve by at most 0.038 points (median 0.012), below the resolution of Fig. 3.

**Re-expression on the Section 3.6 coordinate.** Applying the relation of Appendix A.3 stretches the latent axis of the score-based models by $1/\sqrt{\rho_s} = 1.049$ and leaves the PV-based ones unchanged, raising the PV-ANN − FS-ANN separation over the upper half of the displayed range from $+0.19$ to $+0.26$ points. Against a residual standard deviation of 6.0 both values are below the resolution of Fig. 3, so the figure does not discriminate the two conventions and neither value is read as evidence. Values are produced by `python scripts/convert_pdp_coordinate.py` in the deposited package.

**LMS-Q vertex coordinate.** LMS-Q is constrained to a symmetric quadratic form about its vertex, which the standardized coefficients of Table H2 place at $\eta_1 = -\gamma_1/(2\gamma_{11}) = -0.722/(2 \times 0.364) = -0.99$ on the $z$-score scale. The shallow minimum near it reflects the positive quadratic coefficient, and the apparent asymmetry of the LMS-Q curve within the displayed range arises from the range being off-center relative to that vertex, whereas PV-ANN's asymmetry reflects the unconstrained functional form.

**Upper-tail ordering.** Per-model values at the upper grid points, which Section 5.3 does not interpret, are reproducible via `python scripts/verify_pdp_tail.py` in the deposited package.